\documentclass[twocolumn]{aastex631}
\usepackage{xfrac,graphicx,color,float,hyperref}
\usepackage{soul} 

\newcommand{\bagpipes}{\textsc{bagpipes}}
\newcommand{\sersic}{S\'ersic}
\newcommand{\beagle}{\textsc{beagle}}
\received{August 1, 2023}

\submitjournal{ApJ}

\begin{document}

\title{Unveiling the distant Universe: Characterizing $z\ge9$ Galaxies in the first epoch of COSMOS-Web}

\shortauthors{Franco et al.}
\suppressAffiliations
\correspondingauthor{M. Franco}
\email{maximilien.franco@austin.utexas.edu}


\author[0000-0002-3560-8599]{Maximilien Franco}
\affiliation{The University of Texas at Austin, 2515 Speedway Blvd Stop C1400, Austin, TX 78712, USA}

\author[0000-0003-3596-8794]{Hollis B. Akins}
\affiliation{The University of Texas at Austin, 2515 Speedway Blvd Stop C1400, Austin, TX 78712, USA}

\author[0000-0002-0930-6466]{Caitlin M. Casey}
\affiliation{The University of Texas at Austin, 2515 Speedway Blvd Stop C1400, Austin, TX 78712, USA}

\author[0000-0001-8519-1130]{Steven L. Finkelstein}
\affiliation{The University of Texas at Austin, 2515 Speedway Blvd Stop C1400, Austin, TX 78712, USA}

\author[0000-0002-7087-0701]{Marko Shuntov}
\affiliation{Cosmic Dawn Center (DAWN), Denmark}

\author[0000-0003-4922-0613]{Katherine Chworowsky}
\affiliation{The University of Texas at Austin, 2515 Speedway Blvd Stop C1400, Austin, TX 78712, USA}

\author[0000-0002-9382-9832]{Andreas L. Faisst}
\affiliation{Caltech/IPAC, 1200 E. California Blvd., Pasadena, CA 91125, USA}

\author[0000-0001-7201-5066]{Seiji Fujimoto}\altaffiliation{NASA Hubble Fellow}
\affiliation{The University of Texas at Austin, 2515 Speedway Blvd Stop C1400, Austin, TX 78712, USA}

\author[0000-0002-7303-4397]{Olivier Ilbert}
\affiliation{Aix Marseille Universit\'{e}, CNRS, CNES, LAM, Marseille, France}

\author[0000-0002-6610-2048]{Anton M. Koekemoer}
\affiliation{Space Telescope Science Institute, 3700 San Martin Dr., Baltimore, MD 21218, USA}

\author[0000-0001-9773-7479]{Daizhong Liu}
\affiliation{Max-Planck-Institut f\"ur Extraterrestrische Physik (MPE), Giessenbachstr. 1, D-85748 Garching, Germany}

\author[0000-0001-7964-5933]{Christopher C. Lovell}
\affiliation{Institute of Cosmology and Gravitation, University of Portsmouth, Burnaby Road, Portsmouth, PO1 3FX, UK}
\affiliation{Astronomy Centre, University of Sussex, Falmer, Brighton BN1 9QH, UK}

\author[0000-0001-7711-3677]{Claudia Maraston}
\affiliation{Institute of Cosmology and Gravitation, University of Portsmouth, Dennis Sciama Building, Burnaby Road, Portsmouth, PO13FX, UK}

\author[0000-0002-9489-7765]{Henry Joy McCracken}
\affiliation{Institut d’Astrophysique de Paris, UMR 7095, CNRS, and Sorbonne Université, 98 bis boulevard Arago, F-75014 Paris, France}

\author[0000-0002-6149-8178]{Jed McKinney}
\affiliation{The University of Texas at Austin, 2515 Speedway Blvd Stop C1400, Austin, TX 78712, USA}

\author[0000-0002-4271-0364]{Brant E. Robertson}
\affiliation{Department of Astronomy and Astrophysics, University of California, Santa Cruz, 1156 High Street, Santa Cruz, CA 95064, USA}

\author[0000-0002-9921-9218]{Micaela B. Bagley}
\affiliation{The University of Texas at Austin, 2515 Speedway Blvd Stop C1400, Austin, TX 78712, USA}

\author[0000-0002-6184-9097]{Jaclyn B. Champagne}
\affiliation{Steward Observatory, University of Arizona, 933 N Cherry Ave, Tucson, AZ 85721, USA}

\author[0000-0003-3881-1397]{Olivia R. Cooper}\altaffiliation{NSF Graduate Research Fellow}
\affiliation{The University of Texas at Austin, 2515 Speedway Blvd Stop C1400, Austin, TX 78712, USA}

\author[0000-0001-8917-2148]{Xuheng Ding}
\affiliation{Kavli Institute for the Physics and Mathematics of the Universe (WPI), The University of Tokyo, Kashiwa, Chiba 277-8583, Japan}

\author[0000-0003-4761-2197]{Nicole E. Drakos}
\affiliation{Department of Astronomy and Astrophysics, University of California, Santa Cruz, 1156 High Street, Santa Cruz, CA 95064, USA}

\author[0000-0002-0200-2857]{Andrea Enia}
\affiliation{University of Bologna—Department of Physics and Astronomy ``Augusto Righi'' (DIFA), Via Gobetti 93/2, I-40129, Bologna, Italy}
\affiliation{INAF—Osservatorio di Astrofisica e Scienza dello Spazio, Via Gobetti 93/3, I-40129, Bologna, Italy}

\author[0000-0001-9885-4589]{Steven Gillman}
\affiliation{Cosmic Dawn Center (DAWN), Denmark}
\affiliation{DTU-Space, Technical University of Denmark, Elektrovej 327, DK-2800 Kgs. Lyngby, Denmark}

\author[0000-0002-0236-919X]{Ghassem Gozaliasl}
\affiliation{Department of Physics, University of Helsinki, P.O. Box 64, FI-00014 Helsinki, Finland}
\affiliation{Department of Computer Science, Aalto University, PO Box 15400, Espoo, FI-00 076, Finland}

\author[0000-0003-0129-2079]{Santosh Harish}
\affiliation{Laboratory for Multiwavelength Astrophysics, School of Physics and Astronomy, Rochester Institute of Technology, 84 Lomb Memorial Drive, Rochester, NY 14623, USA}

\author[0000-0003-4073-3236]{Christopher C. Hayward}
\affiliation{Center for Computational Astrophysics, Flatiron Institute, 162 Fifth Avenue, New York, NY 10010, USA}

\author[0000-0002-3301-3321]{Michaela Hirschmann}
\affiliation{Institute of Physics, GalSpec, Ecole Polytechnique Federale de Lausanne, Observatoire de Sauverny, Chemin Pegasi 51, 1290 Versoix, Switzerland}
\affiliation{INAF, Astronomical Observatory of Trieste, Via Tiepolo 11, 34131 Trieste, Italy}

\author[0000-0002-8412-7951]{Shuowen Jin}
\affiliation{Cosmic Dawn Center (DAWN), Denmark}
\affiliation{DTU-Space, Technical University of Denmark, Elektrovej 327, DK-2800 Kgs. Lyngby, Denmark}

\author[0000-0001-9187-3605]{Jeyhan S. Kartaltepe}
\affiliation{Laboratory for Multiwavelength Astrophysics, School of Physics and Astronomy, Rochester Institute of Technology, 84 Lomb Memorial Drive, Rochester, NY 14623, USA}

\author[0000-0002-5588-9156]{Vasily Kokorev}
\affiliation{Kapteyn Astronomical Institute, University of Groningen, PO Box 800, 9700 AV Groningen, The Netherlands}

\author[0009-0008-5926-818X]{Clotilde Laigle}
\affiliation{Institut d’Astrophysique de Paris, UMR 7095, CNRS, and Sorbonne Université, 98 bis boulevard Arago, F-75014 Paris, France}

\author[0000-0002-7530-8857]{Arianna S. Long}\altaffiliation{NASA Hubble Fellow}
\affiliation{The University of Texas at Austin, 2515 Speedway Blvd Stop C1400, Austin, TX 78712, USA}

\author[0000-0002-4872-2294]{Georgios Magdis}
\affiliation{Cosmic Dawn Center (DAWN), Denmark}
\affiliation{DTU-Space, Technical University of Denmark, Elektrovej 327, DK-2800 Kgs. Lyngby, Denmark}
\affiliation{Niels Bohr Institute, University of Copenhagen, Jagtvej 128, DK-2200, Copenhagen, Denmark}

\author[0000-0003-3266-2001]{Guillaume Mahler}
\affil{Department of Physics, Centre for Extragalactic Astronomy, Durham University, South Road, Durham DH1 3LE, UK}
\affil{Department of Physics, Institute for Computational Cosmology, Durham University, South Road, Durham DH1 3LE, UK}

\author[0000-0001-9189-7818]{Crystal L. Martin}
\affil{Department of Physics, University of California, Santa Barbara, Santa Barbara, CA 93109, USA}

\author[0000-0001-5846-4404]{Bahram Mobasher}
\affiliation{Department of Physics and Astronomy, University of California, Riverside, 900 University Avenue, Riverside, CA 92521, USA}

\author[0000-0003-2397-0360]{Louise Paquereau}
\affiliation{Institut d’Astrophysique de Paris, UMR 7095, CNRS, and Sorbonne Université, 98 bis boulevard Arago, F-75014 Paris, France}

\author[0000-0002-7093-7355]{Alvio Renzini}
\affiliation{INAF, Osservatorio Astronomico di Padova, Vicolo dell’Osservatorio 5, I-35122 Padova, Italy}

\author[0000-0002-4485-8549]{Jason Rhodes}
\affiliation{Jet Propulsion Laboratory, California Institute of Technology, 4800 Oak Grove Drive, Pasadena, CA 91001, USA}

\author[0000-0003-0427-8387]{R. Michael Rich}
\affiliation{Department of Physics and Astronomy, UCLA, PAB 430 Portola Plaza, Box 951547, Los Angeles, CA 90095-1547}

\author[0000-0002-5496-4118]{Kartik Sheth}
\affiliation{NASA Headquarters, 300 Hidden Figures Way SE, Washington DC 20546}

\author[0000-0002-0000-6977]{John D. Silverman}
\affiliation{Kavli Institute for the Physics and Mathematics of the Universe (Kavli IPMU, WPI), UTIAS, Tokyo Institutes for Advanced Study, University of Tokyo, Chiba, 277-8583, Japan}
\affiliation{Department of Astronomy, School of Science, The University of Tokyo, 7-3-1 Hongo, Bunkyo, Tokyo 113-0033, Japan}
\affiliation{Center for Data-Driven Discovery, Kavli IPMU (WPI), UTIAS, The University of Tokyo, Kashiwa, Chiba 277-8583, Japan}

\author[0000-0002-9735-3851]{Martin Sparre}
\affiliation{Institut f\"ur Physik und Astronomie, Universit\"at Potsdam, Karl-Liebknecht-Str.\,24/25, 14476 Golm, Germany}

\author[0000-0003-4352-2063]{Margherita Talia}
\affiliation{University of Bologna - Department of Physics and Astronomy “Augusto Righi” (DIFA), Via Gobetti 93/2, I-40129 Bologna, Italy}
\affiliation{INAF, Osservatorio di Astrofisica e Scienza dello Spazio, Via Gobetti 93/3, I-40129, Bologna, Italy}

\author[0000-0002-3683-7297]{Benny Trakhtenbrot}
\affiliation{School of Physics and Astronomy, Tel Aviv University, Tel Aviv 69978, Israel}

\author[0000-0001-6477-4011]{Francesco Valentino}
\affiliation{Cosmic Dawn Center (DAWN), Denmark} 
\affiliation{Niels Bohr Institute, University of Copenhagen, Jagtvej 128, DK-2200, Copenhagen, Denmark}

\author[0000-0002-1905-4194]{Aswin P. Vijayan}
\affiliation{Cosmic Dawn Center (DAWN), Denmark} 
\affiliation{DTU-Space, Technical University of Denmark, Elektrovej 327, DK-2800 Kgs. Lyngby, Denmark}

\author[0000-0003-3903-6935]{Stephen M.~Wilkins}
\affiliation{Astronomy Centre, University of Sussex, Falmer, Brighton BN1 9QH, UK}
\affiliation{Institute of Space Sciences and Astronomy, University of Malta, Msida MSD 2080, Malta}

\author[0000-0002-8434-880X]{Lilan Yang}
\affiliation{Kavli Institute for the Physics and Mathematics of the Universe (WPI), The University of Tokyo, Kashiwa, Chiba 277-8583, Japan}

\author[0000-0002-7051-1100]{Jorge A. Zavala}
\affiliation{National Astronomical Observatory of Japan, 2-21-1 Osawa, Mitaka, Tokyo 181-8588, Japan}
\collaboration{51}{\vspace{-20pt}}





\begin{abstract}
We report the identification of 15 galaxy candidates at $z\ge9$ using the initial COSMOS-Web \textit{JWST} observations over 77 arcmin$^2$ through four NIRCam filters (F115W, F150W, F277W, F444W) with an overlap with MIRI (F770W) of 8.7 arcmin$^2$. We fit the sample using several publicly-available SED fitting and photometric redshift codes and determine their redshifts between $z=9.3$ and $z=10.9$ ($\langle z\rangle=10.0$), UV-magnitudes between M$_{\rm UV}$ = $-$21.2 and $-$19.5 (with $\langle $M$_{\rm UV}\rangle=-20.2$) and rest-frame UV slopes ($\langle \beta\rangle=-2.4$). These galaxies are, on average, more luminous than most $z\ge9$ candidates discovered by \textit{JWST} so far in the literature, while exhibiting similar blue colors in their rest-frame UV. The rest-frame UV slopes derived from SED-fitting are blue ($\beta\sim$[$-$2.0, $-$2.7]) without reaching extremely blue values as reported in other recent studies at these redshifts. The blue color is consistent with models that suggest the underlying stellar population is not yet fully enriched in metals like similarly luminous galaxies in the lower redshift Universe. The derived stellar masses with $\langle \log_{\rm 10} ($M$_\star/$M$_\odot)\rangle\approx8-9$ are not in tension with the standard $\Lambda$CDM model and our measurement of the volume density of such UV luminous galaxies aligns well with previously measured values presented in the literature at $z\sim9-10$. Our sample of galaxies, although compact, are significantly resolved.
\end{abstract}

\section{Introduction} \label{sec:intro}
The search for, and characterisation of, the most distant galaxies in the Universe is fundamental to our understanding of galaxy formation and evolution as a whole. However, we know very little about the first galaxies at $z>10$. Prior to \textit{JWST}, only one galaxy was confirmed at these redshifts (GN-z11; \citealt{oesch16,bunker23}). The recent launch of \textit{JWST}, with its unprecedented angular resolution and sensitivity at infrared wavelengths, is now providing the ideal observations to study star-formation in high-redshift galaxies at rest-frame optical wavelengths. As a result, the number of $z\ge9$ candidates has increased significantly \citep[e.g.,][Casey et al 2023 in prep.]{pontoppidan22, naidu22, whitler23, finkelstein22b,adams23,austin23, leung23} with some spectroscopically-confirmed systems reaching up to z$\sim$13, \citep{robertson23,curtis-lake23} and candidates up to $z \sim$ 17 \citep{harikane22, atek23a, finkelstein23a, austin23, hainline23}.

Extragalactic \textit{JWST} surveys that aim for the high-redshift Universe have adopted various, complementary strategies, including some very deep surveys (e.g., NGDEEP, GO \#2079), others combining advantageously the observations of several \textit{JWST} instruments (JADES, GTO\,\#1180, 1210 \&\ 1287, and CEERS, ERS\,\#1345; \citealt{eisenstein23}) or observing strongly lensed fields (e.g., UNCOVER, GO\,\#\,2561). The unique strength of the COSMOS-Web program (GO \# 1727; \citealt{casey23a}) is its large area, three times larger than all other \textit{JWST} deep field programs combined; when complete, in January 2024, COSMOS-Web will cover a contiguous 0.54 deg$^2$ with the Near Infrared Camera (NIRCam; \citealt{rieke03,rieke05,beichman12,rieke23}).  This wide area opens up a specific parameter space during the epoch of reionization (EoR), which remains inaccessible with other smaller surveys of particularly intrinsically bright galaxies (M$_{\rm UV} < -20$; \citealt{finkelstein23a}) while effectively reducing uncertainties in fundamental extragalactic measurements resulting from cosmic variance (at $z\sim9$ the cosmic variance, $\sigma_v^2$, is less than 10\%; \citealt{trenti08, casey23a}). These particularly distant and bright galaxies are prime candidates to constrain the early growth of structures and galaxy formation and evolution models \citep{finkelstein23a, mason23, yung23}. In addition, this wide area can uniquely probe the different environments in terms of galaxy density and spatial ionization of neutral hydrogen.

The epoch of reionization (finishing around $z\sim6$; \citealt{stanway03}, with a mid-point at $z=7.7\pm0.8$; \citealt{planck20}) marks a crucial period in the history of the Universe, when the first stars and galaxies formed and started to emit ultraviolet radiation that ionized the neutral hydrogen gas in the intergalactic medium. Quantifying key properties of these galaxies, such as their UV magnitude (M$_{\rm UV}$), star formation rates (SFR), stellar masses (M$_*$), and UV beta slope ($\beta$) can provide valuable insights into these processes. Recent studies have highlighted the importance of the host dark matter halos and the sources that ionized the intergalactic medium (IGM) during the epoch of reionization in shaping the properties of the first galaxies \cite[e.g.,][]{hutter21}.

In this paper, we identify 15 new high$-z$ ($z\ge9$) galaxy candidates in the first epoch of COSMOS-Web (4\% of the total survey area), and employ a range of Spectral Energy Distribution (SED) fitting techniques to accurately derive their physical properties in the early Universe. In Section~\ref{sec:data}, we outline the observations used for the detection of these galaxies, and we describe the selection method in Section~\ref{sec:methods}. We present the sample and the results of our SED fitting in Section~\ref{sec:results}, the UV luminosity function in Section~\ref{sec:UVLF} and the implications of these results on our understanding of early galaxy growth evolution in Section~\ref{sec:discussion}. Throughout this paper, we adopt a spatially flat $\Lambda$CDM cosmological model with H$_0$\,=\,70 km\,s$^{-1}$Mpc$^{-1}$, $\Omega_m$\,=\,0.3 and $\Omega_{\Lambda}$\,=\,0.7. We assume a Chabrier \citep{chabrier03} Initial Mass Function (IMF). All magnitudes are quoted in the AB system \citep{oke83}.

\begin{figure*}[t]
  \centering
    \includegraphics[width=0.8\textwidth]{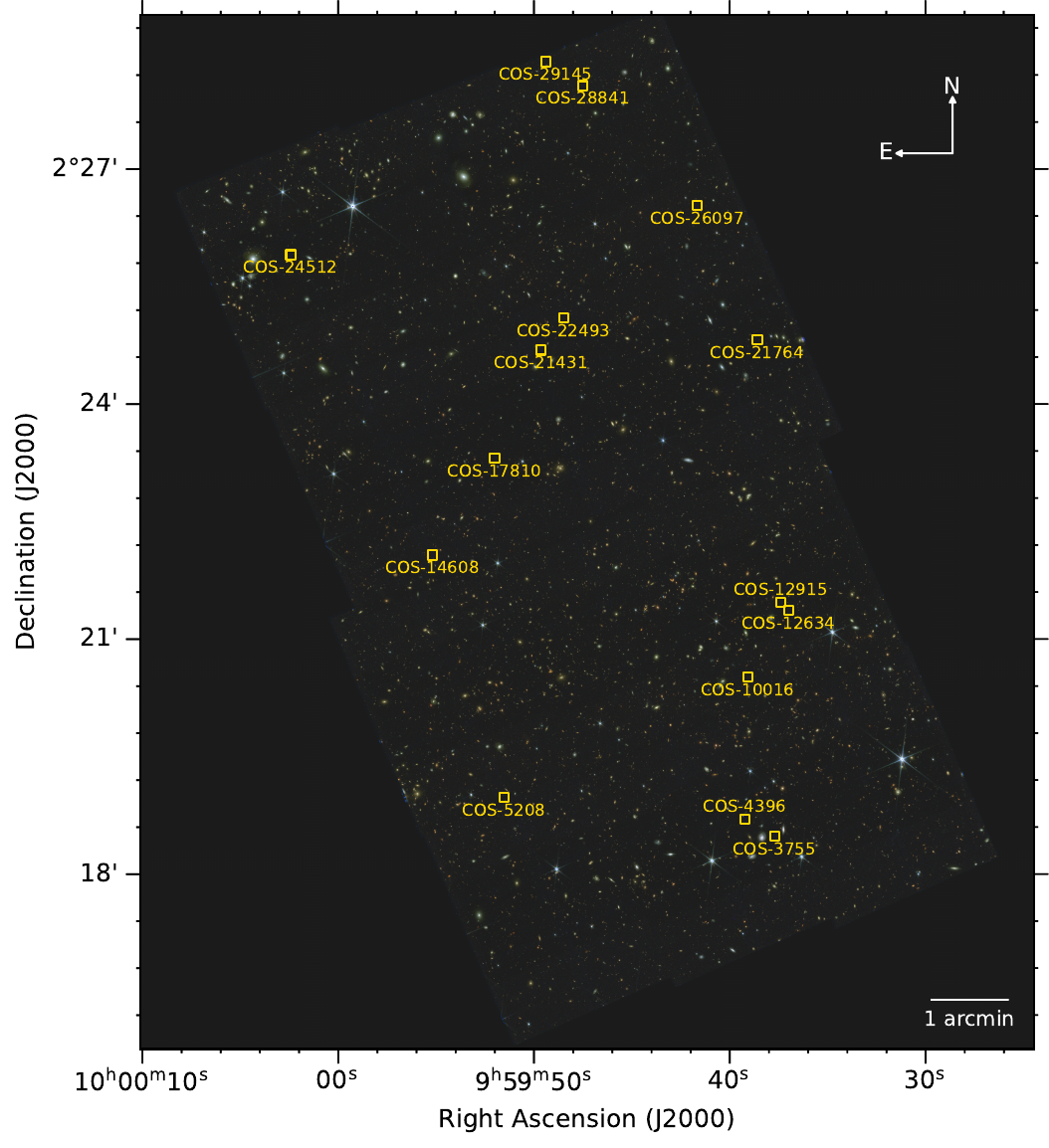}
  \caption{\textit{JWST}/NIRCam color image (F115W, F150W, F277W, F444W) of he first six visits (out of a total of 152) of COSMOS-Web. The positions of our 15 high$-z$ galaxy candidates are indicated by the yellow squares.}
  \label{Fig::position}
\end{figure*}

\section{Observations \& Data Reduction}
\label{sec:data}

\subsection{The COSMOS-Web survey}
The COSMOS-Web survey (\citealt{casey23a}; GO \# 1727) was selected as a 255-hour \textit{JWST} treasury program.  The full survey will map a contiguous 0.54 deg$^2$ area using the Near-Infrared Camera (NIRCam) and 0.19 deg$^2$ using the Mid-Infrared Imager (MIRI) in the COSMOS field \citep{scoville07a,capak07a,koekemoer07a}. It is the largest \textit{JWST} program both in terms of area covered at this depth and GO time allocated. The depths of the NIRCam data are measured to be 26.6-27.3 AB (F115W), 26.9-27.7 (F150W), 27.5-28.2 (F277W), and 27.5-28.2 (F444W) for 5$\sigma$ point sources calculated within 0.15" radius apertures. For the MIRI F770W filter, the depths calculated within 0"3 radius apertures vary between 25.33 and 25.98 AB for point sources. The variable depth is caused by different portions of the mosaic being covered with a different number of exposures, as the mosaicing of COSMOS-Web uses the somewhat large 4TIGHT dither pattern.  More details about the design and motivation for the COSMOS-Web survey are given in \citet{casey23a}.

The first 4\% of data (six of 152 visits) for COSMOS-Web were taken in early January 2023 and cover a total area of $\sim$77 arcmin$^2$ with NIRCam ($\sim$8.7\,arcmin$^2$ of which is also covered by the MIRI parallels).  This paper focuses on sources discovered in this initial imaging area. 

 \begin{deluxetable*}{lccccccc}
 \tabletypesize{\scriptsize}
 \tablecaption{Photometry of the $z\ge9$ galaxy candidate sample. \label{tab:photometry}}
  \tablehead{
  \colhead{ID} &  \colhead{RA} & \colhead{Dec} & \colhead{\textit{JWST}/} & \colhead{\textit{JWST}/} & \colhead{\textit{JWST}/} & \colhead{\textit{JWST}/} & \colhead{\textit{JWST}/}  \\[-0.4cm]
  \colhead{} & \colhead{J2000} & \colhead{J2000} & \colhead{F115W} & \colhead{F150W} & \colhead{F277W} & \colhead{F444W} & \colhead{F770W}}
 \colnumbers
 \startdata
          COS-28841 & 149.94791615 & 2.46773220 & $>$27.68 & 26.64 $\pm$ 0.11 & 26.79 $\pm$ 0.07 & 26.65 $\pm$ 0.07 &  ...      \\ 
          COS-17810 & 149.96667829 & 2.38852142 & $>$28.00 & 27.10 $\pm$ 0.12 & 26.93 $\pm$ 0.06 & 26.94 $\pm$ 0.07 &  ...      \\ 
          COS-29145 & 149.95579364 & 2.47287402 & $>$27.68 & 26.88 $\pm$ 0.13 & 27.18 $\pm$ 0.10 & 26.82 $\pm$ 0.08 &  ...      \\ 
          COS-5208  & 149.96468292 & 2.31632995 & $>$27.81 & 27.06 $\pm$ 0.14 & 27.68 $\pm$ 0.14 & 26.74 $\pm$ 0.09 &  ...      \\ 
          COS-12915 & 149.90581503 & 2.35780491 & $>$28.00 & 27.02 $\pm$ 0.11 & 27.47 $\pm$ 0.10 & 27.70 $\pm$ 0.13 &  ...      \\ 
          COS-21764 & 149.91077923 & 2.41372492 & $>$28.00 & 27.16 $\pm$ 0.13 & 27.78 $\pm$ 0.13 & 27.16 $\pm$ 0.08 &  ...      \\ 
          COS-26097 & 149.92357696 & 2.44227575 & $>$27.68 & 27.24 $\pm$ 0.18 & 27.53 $\pm$ 0.13 & 27.58 $\pm$ 0.16 &  ...      \\ 
          COS-22493 & 149.95203282 & 2.41836544 & $>$27.68 & 27.36 $\pm$ 0.20 & 27.43 $\pm$ 0.12 & 27.19 $\pm$ 0.11 &  ...      \\ 
          COS-12634 & 149.90414975 & 2.35610310 & $>$27.68 & 27.42 $\pm$ 0.21 & 27.62 $\pm$ 0.14 & 27.77 $\pm$ 0.19 &  ...      \\ 
          COS-14608 & 149.97992311 & 2.36788542 & $>$27.42 & 27.30 $\pm$ 0.23 & 27.50 $\pm$ 0.17 & 26.99 $\pm$ 0.12 &  ...      \\ 
          COS-4396  & 149.91345122 & 2.31169605 & $>$28.00 & 27.40 $\pm$ 0.16 & 27.83 $\pm$ 0.13 & 27.20 $\pm$ 0.09 & $>$26.25 \\ 
          COS-24512 & 150.01008463 & 2.43176823 & $>$27.68 & 27.17 $\pm$ 0.17 & 28.04 $\pm$ 0.21 & 28.29 $\pm$ 0.29 &  ...      \\ 
          COS-10016 & 149.91281685 & 2.34193317 & $>$27.68 & 27.41 $\pm$ 0.21 & 27.97 $\pm$ 0.19 & 27.61 $\pm$ 0.16 &    ...    \\ 
          COS-21431 & 149.95684396 & 2.41151565 & $>$27.68 & 27.65 $\pm$ 0.26 & 28.30 $\pm$ 0.25 & 27.43 $\pm$ 0.14 &    ...    \\ 
          COS-3755  & 149.90714647 & 2.30812336 & $>$28.00 & 27.89 $\pm$ 0.24 & 28.34 $\pm$ 0.21 & 28.32 $\pm$ 0.22 & $>$26.53 \\ 
\enddata
\tablecomments{Coordinates and multi-band photometry of our galaxy sample with fluxes obtained from \texttt{SE++} model-based photometry. No fluxes (above 2$\sigma$) have been detected in filters blueward of the supposed Ly$\alpha$ break. VISTA $YJHK_s$ are not constraining for these sources, we have omitted them when fitting.}
\end{deluxetable*}

\subsubsection{NIRCam}
As part of COSMOS-Web, observations were taken through the four NIRCam wide-band filters: F115W, F150W, F277W, and F444W. The full data reduction will be described in detail in M. Franco et al. in prep., with the main steps summarized here.
After retrieving all the uncalibrated NIRCam images from the STScI MAST Archive\footnote{\url{https://archive.stsci.edu/}}, the images were reduced using the \textit{JWST} Calibration Pipeline\footnote{\url{https://github.com/spacetelescope/jwst}} version 1.8.3 \citep{2023zndo...7714020B}, with the addition of several custom modifications, as has also been done for other \textit{JWST} studies \citep[e.g.,][]{finkelstein22b, bagley22a}, including correcting the 1/f noise and subtraction of low-level background. We used the Calibration Reference Data System (CRDS)\footnote{\url{https://jwst-crds.stsci.edu}} pmap 0989 which corresponds to the NIRCam instrument mapping imap 0232, where some reference files include in-flight data, and which represented the most current calibrations when our observations were obtained.

The final mosaics are created in Stage 3 of the pipeline which vary only in resolution with a pixel size of 0.03"/pixel and 0.06"/pixel. Unless otherwise stated, we will use the 0.03"/pixel resolution mosaic in the following.

Achieving precise absolute and relative astrometry of the \textit{JWST} mosaics across all the filters is essential for ensuring the reliability of the resulting measurements, including photometry, morphology, and photometric redshift determination. Our astrometric calibration is carried out by utilizing the \textit{JWST} TweakReg procedure which is part of the \textit{JWST} pipeline. To carry out this step, we first generated a reference catalog over our COSMOS-Web region using a new 0.03"/pixel mosaic of the original COSMOS \textit{HST}/F814W imaging data, which had been reprocessed following the methodology described in \cite{koekemoer11}, in particular with improved astrometric alignment to Gaia-DR3 \citep{gaia21} and the COSMOS2020 catalog \citep{weaver22a}. Our NIRCam data are aligned to this reference catalog with median offsets in RA and Dec below the level of 5 mas, regardless of the filter used, and the median absolute deviation (MAD) values are less than 12 mas across the entire field, with minor variation between the different filters (more details will be provided in Franco et al., in prep.).

\subsubsection{MIRI}
 The MIRI parallels taken in conjunction with NIRCam imaging in January 2023 cover a total area of 27\,arcmin$^2$, and to date 8.7\,arcmin$^2$ of that coverage directly overlaps with the NIRCam imaging.  MIRI data were processed through the \textit{JWST} Calibration Pipeline version 1.8.3, with a two-step procedure (see Harish et al. in prep.). In the first step, we process the MIRI data through stages 1-3 of the \textit{JWST} pipeline, with inflight calibrations applied, and obtain the drizzled mosaic image. Then we detect sources in the mosaic image using \texttt{SExtractor} Classic \citep{bertin96} and build a source emission catalog. We mask out pixels of the source emission in the ``rate" observations (source-emission-masked rate, or ``dark rate"), then for each data set we build a master background rate image by combining dark rate images of other data sets with the same filter and close dates. In the second step, we reprocess the ``rate" image of each data set by using the corresponding master background rate image as the background exposure in stage 2 of the \textit{JWST} pipeline. Then stage 3 of the \textit{JWST} pipeline produces our final drizzled mosaic, with astrometry aligned to that of the new \textit{HST} and COSMOS2020 catalogs (same as for NIRCam).

\subsection{Complementary Data}
In addition to the \textit{JWST} data described above, we take advantage of the rich multiwavelength data available across the COSMOS field.  In this paper, we make use of the {\it Hubble Space Telescope} imaging \citep{koekemoer07a}, consisting of ACS F814W imaging to an average $\sim$5$\sigma$ point source depth of 27.2\, AB mag (in a 0.24" diameter aperture).  We also use the wealth of ground-based optical/near-infrared (OIR) imaging data including Subaru SuprimeCam and Hyper Suprime-Cam imaging \citep{aihara22a} as well as UltraVISTA \citep{mccracken12a} near-infrared imaging data release \#5 (DR5).  The details of the ancillary ground-based imaging are described in detail in \citet{weaver22a}, though they make use of UltraVISTA data release \#4 (DR4) rather than DR5.  In addition to the extensive OIR data in the field, we also use long-wavelength data (submillimeter through radio) as well as X-ray to check for possible emission around the newly-identified sources in this work; these datasets are further described in \citet{casey23a}.

\section{Methods}
\label{sec:methods}

\subsection{Photometry and SED fitting}
\label{sec:SEDfitting}

Our methodology involves utilizing \texttt{SourceXtractor++} (\texttt{SE++}; \citealt{bertin20,kummel20}), an updated version of the widely used \texttt{SExtractor} package \citep{bertin96}, to conduct source detection, model-based photometry, and generate multi-band catalogs. 
We choose to use \texttt{SE++} model-based photometry in order to take full advantage of the depth and filter coverage of seeing-limited ground-based data in COSMOS and high-resolution near-infrared \textit{JWST} imaging. To detect sources, we construct a $\chi^2$ detection image from all four NIRCam bands and use priors for the source centroid positions derived from this image. 
For each detected source in the $\chi^2$ image, \texttt{SE++} then fits a 2D \sersic\ model convolved with the filter-specific PSF in each of the measurement bands (see Shuntov et al. in prep.) using a PSF from \texttt{WebbPSF} \citep{perrin14}.

We note that, even for sources that are not detected in a given band, \texttt{SE++} always fits a model, though the model will be below the noise. In this case, computing photometric uncertainties is a non-trivial task, and errors in the dropout bands can sometimes be significantly underestimated. To address this issue, we set a noise floor for each band that corresponds to the root-mean-square (rms) measured in circular apertures with radii of 0.15” (for ACS/NIRCam), 0.3” (for MIRI), and 1” (for ground-based data). We adopt the measured depths from \cite{weaver22a} and \cite{casey23a}. The full explanation of the catalog creation will be provided in detail in an upcoming paper (Shuntov et al. in prep.). The IDs of the high-redshift galaxy candidates and the fluxes are given in Table~\ref{tab:photometry}. Although each of these fits incorporates slightly different physical assumptions, the diverse range of approaches employed in this study provides a valuable means of testing the validity of the candidate galaxies at $z\ge9$.

\begin{figure*}[h!]
\centering
\includegraphics[width=1\textwidth]{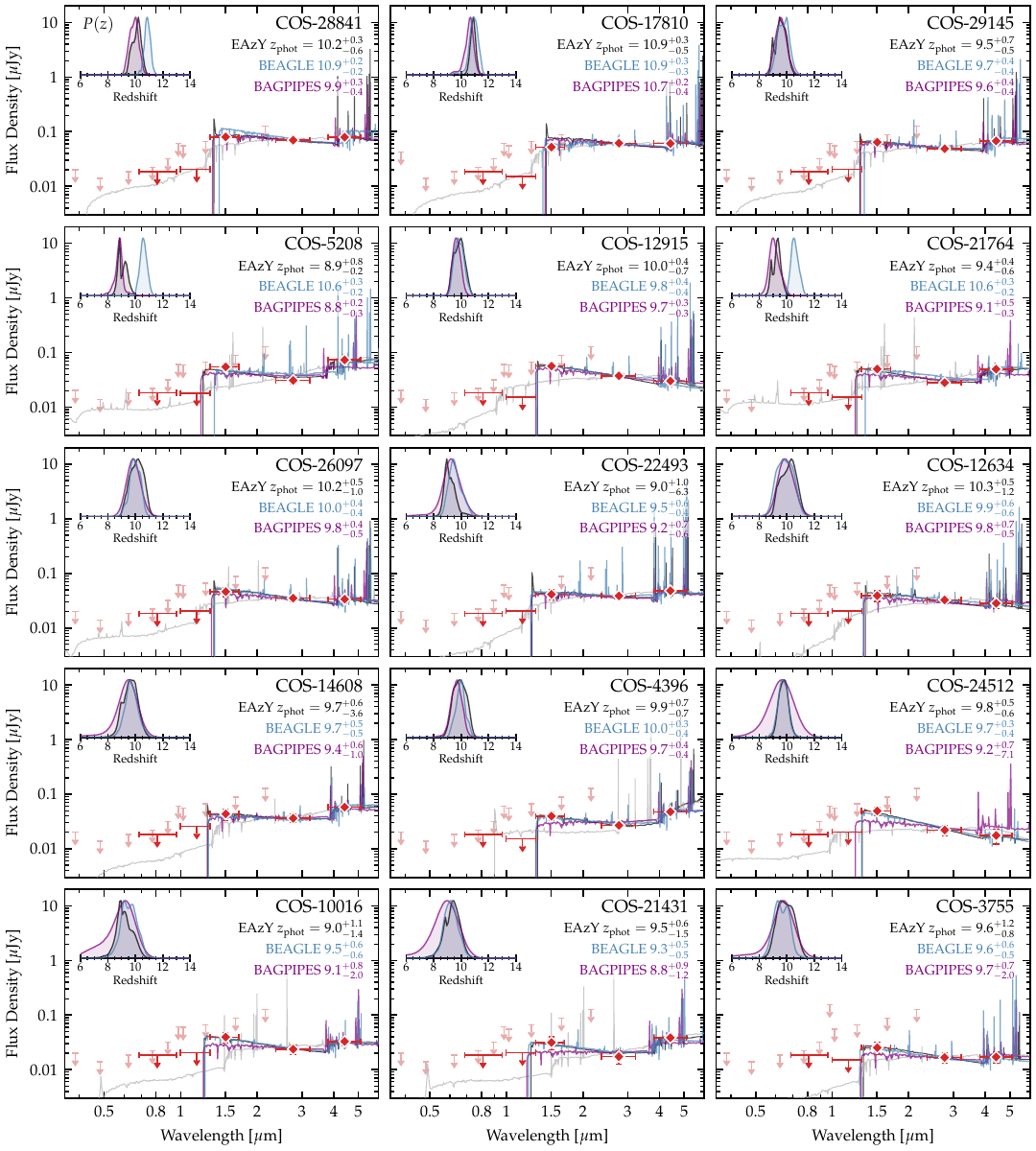} 
\caption{Main panels: Spectral Energy Distributions ordered by $M_{\rm UV}$ (bright-to-faint). Redshift probability distributions are shown in the top left on each panel, and the redshifts themselves are given in the top right, for each SED-fitting code (dark grey: \textsc{EAzY}, blue: \beagle, purple: \bagpipes). Best-fitting SED model templates for each code are also shown with the same color-coding (we add in addition in light grey the low$-z$ solution ($z<7$) from \textsc{EAzY}). Red points, and 2$\sigma$ limits, are observed data.}
         \label{Fig::SEDs}
\end{figure*}

We use several different SED fitting techniques to derive the redshifts and the properties of our galaxies. We use the \textsc{EAzY} \citep{brammer08}, \bagpipes\  \citep{Carnall18} and \beagle\  \citep{gutkin16,chevallard16} SED fitting tools, with the parameters given in the following sections. The different best-fit SEDs are presented in Fig~\ref{Fig::SEDs}. We compare the results coming from these different approaches in Section~\ref{sec:results}.

 \begin{deluxetable*}{lrrrrrr}
 \label{tab:redshift}
 \tabletypesize{\scriptsize}
  \tablewidth{0pt} 
 \tablecaption{Measurements of redshifts}
\tablehead{\colhead{ID}  & \colhead{$z_{\textsc{EAzY}}$} & \colhead{$\chi^2_{\textsc{EAzY}}$}  & \colhead{$\chi^2_{\textsc{EAzY}, \rm low-z}$} & \colhead{$\Delta \chi^2_{\textsc{EAzY}}$} & \colhead{$z_{\textsc{Beagle}}$}   & \colhead{$z_{\textsc{Bagpipes}}$}}
 \colnumbers
\startdata
COS-28841 &  10.2$_{ -0.6}^{+  0.3}$ &   4.6 &   21.1 &  16.5 &  10.9$_{ -0.2}^{+  0.2}$ &   9.9$_{ -0.4}^{+  0.3}$ \\ 
COS-17810 &  10.9$_{ -0.5}^{+  0.3}$ &   9.1 &   13.4 &   4.3 &  10.9$_{ -0.3}^{+  0.3}$ &  10.6$_{ -0.3}^{+  0.3}$ \\ 
COS-29145 &   9.5$_{ -0.5}^{+  0.7}$ &   5.4 &   14.7 &   9.4 &   9.7$_{ -0.4}^{+  0.4}$ &   9.6$_{ -0.4}^{+  0.4}$ \\ 
COS-5208  &   8.9$_{ -0.2}^{+  0.8}$ &   9.9 &   24.9 &  14.9 &  10.6$_{ -0.3}^{+  0.2}$ &   8.8$_{ -0.3}^{+  0.2}$ \\ 
COS-12915 &  10.0$_{ -0.7}^{+  0.4}$ &   4.1 &   24.3 &  20.3 &   9.8$_{ -0.4}^{+  0.4}$ &   9.7$_{ -0.3}^{+  0.3}$ \\ 
COS-21764 &   9.4$_{ -0.6}^{+  0.4}$ &  13.5 &   21.2 &   7.7 &  10.6$_{ -0.3}^{+  0.2}$ &   9.1$_{ -0.3}^{+  0.5}$ \\ 
COS-26097 &  10.2$_{ -1.0}^{+  0.5}$ &   3.9 &    8.2 &   4.4 &  10.0$_{ -0.4}^{+  0.4}$ &   9.8$_{ -0.5}^{+  0.4}$ \\ 
COS-22493 &   9.0$_{ -6.3}^{+  1.0}$ &   4.2 &    6.9 &   2.7 &   9.5$_{ -0.6}^{+  0.4}$ &   9.2$_{ -0.6}^{+  0.7}$ \\ 
COS-12634 &  10.3$_{ -1.2}^{+  0.5}$ &   1.6 &    6.4 &   4.9 &   9.9$_{ -0.6}^{+  0.6}$ &   9.8$_{ -0.5}^{+  0.7}$ \\ 
COS-14608 &   9.7$_{ -3.6}^{+  0.6}$ &   4.3 &    6.6 &   2.3 &   9.7$_{ -0.5}^{+  0.5}$ &   9.4$_{ -1.1}^{+  0.6}$ \\ 
COS-4396  &   9.9$_{ -0.7}^{+  0.7}$ &   1.0 &   16.1 &  15.1 &  10.0$_{ -0.4}^{+  0.3}$ &   9.7$_{ -0.4}^{+  0.4}$ \\ 
COS-24512 &   9.8$_{ -0.6}^{+  0.5}$ &   6.1 &   12.4 &   6.3 &   9.7$_{ -0.4}^{+  0.3}$ &   9.2$_{ -7.1}^{+  0.7}$ \\ 
COS-10016 &   9.0$_{ -1.4}^{+  1.1}$ &   3.2 &    6.6 &   3.4 &   9.5$_{ -0.6}^{+  0.6}$ &   9.1$_{ -2.0}^{+  0.8}$ \\ 
COS-21431 &   9.5$_{ -1.5}^{+  0.6}$ &   4.8 &    8.2 &   3.3 &   9.3$_{ -0.5}^{+  0.5}$ &   8.8$_{ -1.2}^{+  0.9}$ \\ 
COS-3755  &   9.6$_{ -0.8}^{+  1.2}$ &   0.9 &    5.2 &   4.3 &   9.6$_{ -0.6}^{+  0.5}$ &   9.7$_{ -2.0}^{+  0.7}$ \\ 
\enddata
\tablecomments{(1) ID, (2) photometric redshifts from \textsc{EAzY},  (3) and (4) are the $\chi^{2}$  values from \textsc{EAzY} with a redshift between 0 and 15 and \textsc{EAzY} with a redshift range set to be $z<7$, called the ``low-$z$" solution, respectively, (5) gives the difference in $\chi^{2}$ between these two \textsc{EAzY} runs, (6) and (7) are the best redshift solutions from \beagle\  and \bagpipes\  respectively.}
\end{deluxetable*}

\subsubsection{\textsc{EAzY}}
\label{subsec:Eazy}
To compute photometric redshifts to initially select our galaxies, we use the SED-fitting tool \textsc{EAzY}. For this purpose, we use the default Flexible Stellar Population Synthesis (FSPS) templates (specifically, the fsps QSF 12 v3 version; \citealt{conroy10}), and supplement them with six additional templates from \cite{larson22} optimized for selecting galaxies at $z>8$ with \textit{JWST}. These templates are more effective in replicating the blue UV slopes exhibited by galaxies with high redshifts (see Fig.~4 in \citealt{larson22}). They have been created by integrating stellar population spectra from BPASS \citep{eldridge09} with the possibility of incorporating nebular emission data obtained from Cloudy \citep{ferland17} with a high ionization parameter ($\log_{10}$($U$) = -2), low gas-phase metallicities (Z = 0.05 Z$_\odot$), and excluding Ly$\alpha$ emission. We assume a flat redshift (and magnitude) prior and we extend the redshift search between 0.01 and 15 with steps of $\Delta z$=0.01. 
As is common practice in the literature \citep[e.g.,][]{harikane22, finkelstein23a} we additionally perform an \textsc{EAzY} run with a maximum redshift of $z=7$ to compare the best chi-squared between the low- and high-redshift runs.

\subsubsection{\bagpipes}
\label{subsec:Bagpipes}
In order to assess the results for photo-$z$ with an alternative procedure, we employ an alternative SED-fitting tool \bagpipes. Our SED fitting is carried out using a delayed exponentially declining SFH model, where the star formation rate (SFR) follows a functional form of SFR($t$) $\propto$ $t$ exp (-$t/\tau$ ). We slightly modify the publicly-available \bagpipes\ code to parametrize the age of the delayed-$\tau$ SFH as a fraction of the Hubble time at redshift $z$, rather than an absolute age in Gyr. 
This is necessary to provide equal weight to both old and young stellar populations while maintaining a uniform redshift prior.
We assume a Calzetti law \citep{calzetti20} for dust attenuation and BC03 \citep{bruzual2003} models. Absolute attenuation in the V band can vary between 0 and 3 magnitudes. The ionization parameter can vary between $10^{-3}$ and $10^{-1}$, the stellar mass formed between 10$^{7}$M$_\odot$ and 10$^{11}$M$_\odot$, and the metallicity between 0.001 and 1 times the solar metallicity. We have incorporated nebular emission into our study by utilizing the updated Cloudy models.

\subsubsection{\beagle}
\label{subsec:Beagle}
We also utilized the Bayesian tool \beagle\  developed by \cite{chevallard16} to conduct an additional SED fitting. The templates used by \beagle\  were created by \cite{gutkin16} and are based on the 2016 updated version of the BC03 stellar population models and nebular emission calculated using the Cloudy photoionization code \citep{ferland17}. 
Following others in the literature \citep[e.g.,][]{whitler23}, we adopted a constant SFH model and log-uniform priors on the total stellar mass ($\log_{\rm 10} (M_\star/{\rm M}_\odot$) from 5 to 10), the maximum stellar age ($\log_{\rm 10} (t/{\rm yr})$ from 7 to 10) and the stellar metallicity ($\log_{\rm 10} (Z/Z_\odot)$ from $-$2.2 to $-$0.3).
We include dust attenuation following an SMC law with $\tau_V$ varying with a log-uniform prior from 0.001 to 5. 
Finally, we include nebular emission with ionization parameter $\log_{\rm 10} (U)$ varying from $-$4 to $-$1. 
Crucially, we allow a variable Lyman continuum escape fraction $f_{\rm esc}$. 
This allows for young stellar populations with minimal nebular continuum, extending the parameter space covered by the models to bluer UV slopes.

\begin{figure*}
\centering
\begin{minipage}[t]{1.\textwidth}
\resizebox{\hsize}{!} { 
\includegraphics[width=3cm,clip]{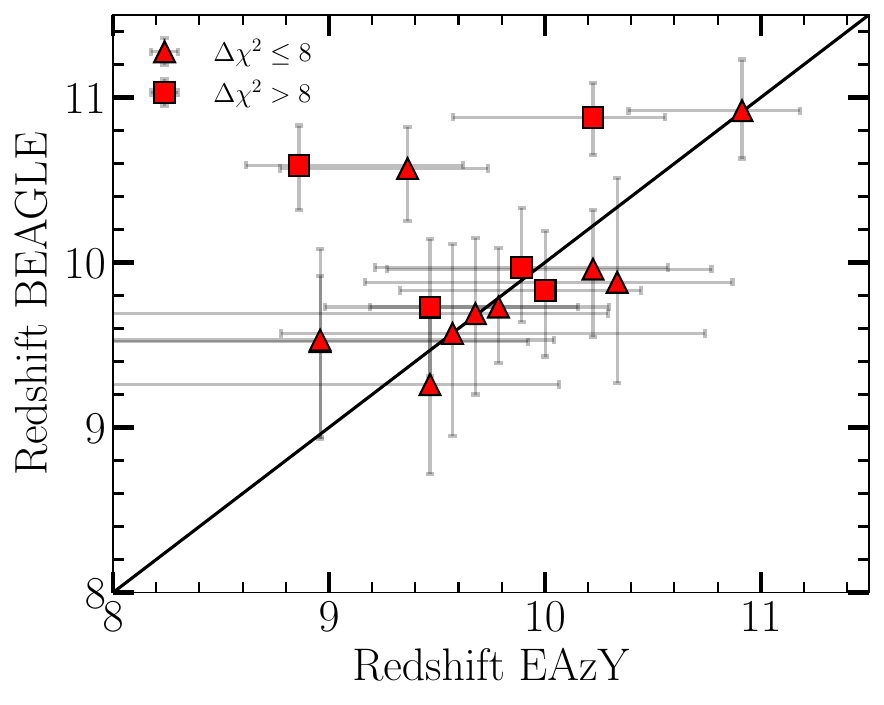} 
\includegraphics[width=3cm,clip]{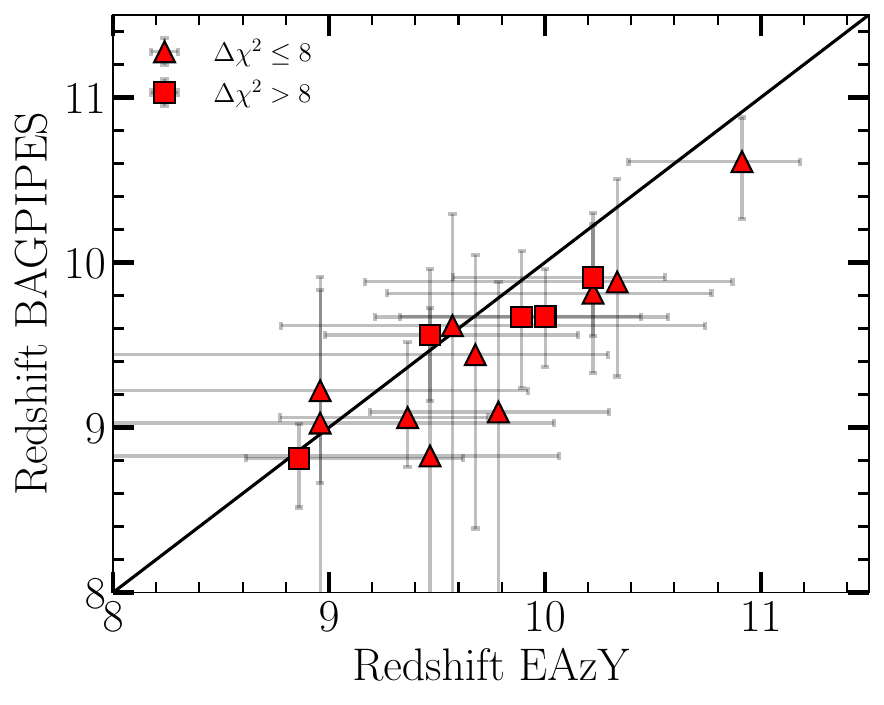} 
\includegraphics[width=3cm,clip]{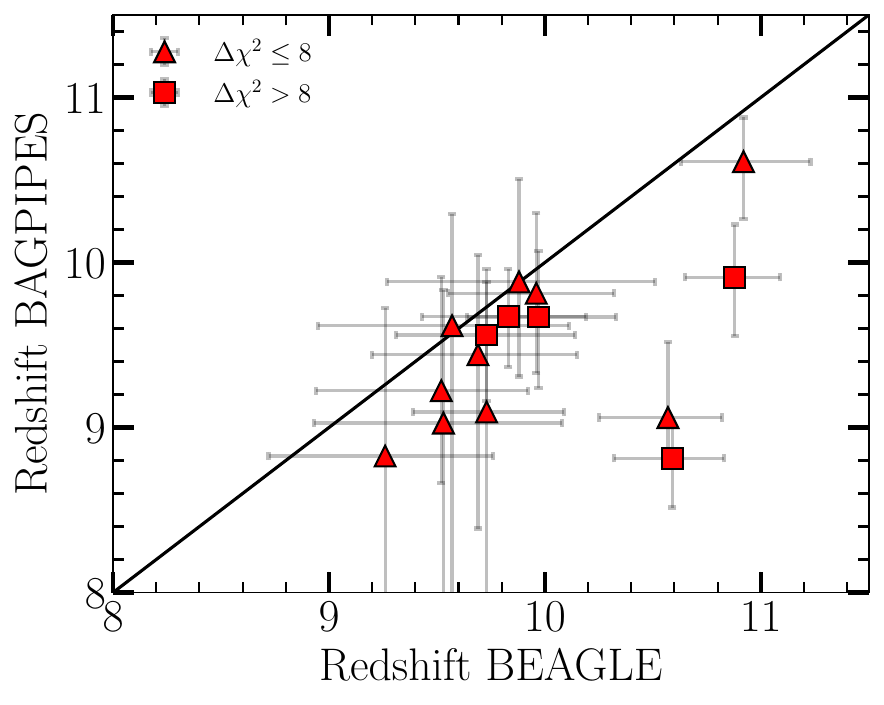} 
}
\end{minipage}
      \caption{Comparison of the redshift obtained with \textsc{EAzY}, \bagpipes, and \beagle. Differences in the model assumptions, priors, and fitting procedures lead to a difference $\Delta z$ /(1+$z_{\rm mean}$) 0.02 between \bagpipes\  and \textsc{EAzY}, 0.05 between \beagle\  and \bagpipes\  and 0.02 between \beagle\  and \textsc{EAzY}. We can however note that \bagpipes\  gives a globally lower redshift than the two other codes. We have differentiated our sample in two sub-samples according to the \textsc{EAzY} $\Delta \chi^2$ between the high redshift solution (unconstrained) and the low redshift solution ($z<7$) with triangle and square markers. No significant differences have been found between these two sub-samples.}
         \label{Fig::Redshift}
\end{figure*}

\subsubsection{Dense-Basis}
We further perform SED fitting for each source using the Dense-Basis SED fitting code \citep{iyer17} to explore the effects of a non-parametric star-formation history on the recovered physical parameters. Dense-Basis uses a flexible star formation history (SFH) represented by a Gaussian Mixture Model \citep{iyer19}. For this work, we define 3 ``shape'' parameters that describe the SFH: $t_{25}, t_{50},$ and $t_{75}$ (requiring the recovered SFH of the galaxy to form ``$x$'' fraction of its total mass by time $t_x$). We impose a uniform (flat) prior on the specific star formation rate (sSFR) with limits on the sSFR (sSFR$/ \textrm{yr}^{-1} \in [10^{-14}, 10^{-7}]$), an exponential prior on the dust attenuation over a wide range of values ($A_V \in [0, 4]$), and a uniform (in log-space) prior on the metallicity ($Z/Z_\odot \in [0.01, 2.0]$). All sources were fit with Dense-Basis assuming a Calzetti dust attenuation law and a Chabrier IMF. We further constrain the redshift range to within 1$\sigma$ of the \beagle\ best-fit redshift to generate posteriors on the galaxies' physical properties under the assumption that redshifts are accurately recovered by \beagle. We use this method to derive the stellar mass and the SFR.

\subsection{Selection of $z\ge9$ Galaxy Candidates}
\label{sec:selection}
In this paper, we focus on the detection of galaxies with $z\ge9$.  At these redshifts, the Ly$\alpha$ break is certain to lie in one of the \textit{JWST} filters, starting from F115W for a $z=9$ galaxy, and shifting to redder filters for higher redshift galaxies. This allows us to identify the drop in flux due to the Ly$\alpha$ break using only NIRCam filters, rather than requiring another instrument such as \textit{HST} (as would be the case for lower-redshift galaxies with a Ly$\alpha$ break at shorter wavelengths). Due to the high sensitivity of NIRCam, this allows us to both reliably identify the presence of a break (thus constructing a robust sample), and measure the $\Delta$(magnitude) across the break that is used to derive galaxy physical properties. This selection method is particularly advantageous for the reliable selection of $z\ge9$ galaxies. 

We construct our sample of high-redshift candidates inspired by the method described in \cite{finkelstein23a}. We employ the following criteria. We require:

\begin{enumerate}
    \item A best-fit photometric redshift ($z_{a}) > 8.9$ from \textsc{EAzY};
    \item A robust detection in the two long wavelength filters. We impose a S/N $>5$ measured in 0.2" diameter apertures in F277W and F444W;
    \item A S/N $<$ 2 in bands blueward of the supposed Ly$\alpha$ break;
    \item $\int \mathcal{P}(z<9) $d$z \leq 0.3$ with \textsc{EAzY}. This means that more than 70\% of the integrated probability is above $z\ge9$ ;
    \item A goodness-of-fit $\chi^2$/N$_{\rm filt} < 3$ with \textsc{EAzY}, where N$_{\rm filt} =5 $, the number of filters that effectively constrain the redshift measurement, here F814W, F115W, F150W, F277W, F444W;
    \item A significantly better fit for the entire redshift range than for redshift restricted between 0 and 7 defined as $\Delta \chi^2 > 2$;
    \item A radius (determined by \texttt{SE++}) greater than 0.01" to remove bad or hot pixels.
    
\end{enumerate}

This selection is intentionally restrictive but does not require extensive visual inspection after these filters have been applied (though all sources that fulfill these criteria were visually inspected). This serves as a pilot study of $z\ge9$ sources in COSMOS-Web and will be superseded with a larger scale study when all survey data is in hand.
Our selection criteria results in a total of 15 $z\ge9$ candidates. We have differentiated in the rest of this study the galaxies for which the redshift is the most reliable with a $\Delta \chi^2$ between the high redshift and low redshift solutions of \textsc{EAzY} greater than 8 (5 galaxies) compared with galaxies with $\Delta \chi^2 \le 8$ (10 galaxies). We advocate for the use of this value to enable direct comparisons between surveys (e.g., CEERS; \citealt{finkelstein23a}). Note that the quantity $\chi^2$ is not reduced.  We double-checked that the sources were not previously detected with ALMA using the A$^3$COSMOS catalog (data version: 20200310; \citealt{liu19}) nor with VLA at 3GHz \citep{smolcic17} within a radius of 0.9".

 \begin{deluxetable*}{lccccccccc}
 \tabletypesize{\scriptsize}
 \tablecaption{Physical properties of the $z>9$ galaxy sample. \label{tab:properties}}
 \tablehead{
 \colhead{ID} & \colhead{M$_{\rm UV}$} & \colhead{$\beta$} & \colhead{$\log_{10}$(M$_\star$/M$_{\odot})$}  & \colhead{SFR$_{\rm 10}$} & \colhead{A$_V$}  & \colhead{R$_{\rm eff}$} & \colhead{R$_{\rm eff}$} & \colhead{n} & \colhead{age$_{50}$} \\[-0.25cm]
 \colhead{} &  \colhead{[mag]} & \colhead{} & \colhead{M$_{\odot}$} & \colhead{[M$_{\odot}$ yr$^{-1}$]} & \colhead{[mag]} & \colhead{[mas]} & \colhead{[pc]} & \colhead{}  & \colhead{Gyr}
 }
\colnumbers
\startdata
          COS-28841  & $-21.21_{-0.11}^{+0.11}$ & $-2.54_{-0.13}^{+0.17}$ & $8.8_{-0.3}^{+0.3}$ & $13.1_{-2.8}^{+2.5}$ & $0.01_{-0.01}^{+0.04}$ & 111 $\pm$  8 & 437 $\pm$  33 & 0.5 $\pm$ 0.1 & $0.05_{- 0.04}^{+ 0.17}$ \\
          COS-17810  & $-20.94_{-0.13}^{+0.19}$ & $-2.42_{-0.17}^{+0.27}$ & $8.9_{-0.4}^{+0.5}$ & $ 8.7_{-7.0}^{+7.2}$ & $0.05_{-0.04}^{+0.08}$ & 137 $\pm$  7 & 536 $\pm$  29 & 1.0 $\pm$ 0.1 & $0.15_{- 0.14}^{+ 0.09}$ \\
          COS-29145  & $-20.52_{-0.16}^{+0.14}$ & $-2.33_{-0.19}^{+0.25}$ & $9.5_{-0.5}^{+0.2}$ & $ 8.0_{-3.8}^{+3.0}$ & $0.04_{-0.03}^{+0.07}$ &  61 $\pm$ 12 & 262 $\pm$  52 & 1.9 $\pm$ 0.6 & $0.26_{- 0.11}^{+ 0.09}$ \\
          COS-5208  & $-20.40_{-0.11}^{+0.13}$ & $-2.19_{-0.13}^{+0.18}$ & $8.8_{-0.5}^{+0.5}$ & $ 6.2_{-1.6}^{+1.9}$ & $0.03_{-0.02}^{+0.05}$ & 130 $\pm$ 23 & 522 $\pm$  94 & 4.5 $\pm$ 1.2 & $0.16_{- 0.14}^{+ 0.10}$ \\
          COS-12915  & $-20.40_{-0.13}^{+0.12}$ & $-2.67_{-0.12}^{+0.16}$ & $8.4_{-0.3}^{+0.3}$ & $ 5.1_{-1.1}^{+1.2}$ & $0.02_{-0.02}^{+0.04}$ &  96 $\pm$ 12 & 406 $\pm$  51 & 0.7 $\pm$ 0.2 & $0.04_{- 0.03}^{+ 0.21}$ \\
          COS-21764  & $-20.29_{-0.14}^{+0.13}$ & $-2.32_{-0.22}^{+0.15}$ & $8.5_{-0.3}^{+0.5}$ & $ 5.8_{-1.1}^{+1.1}$ & $0.02_{-0.01}^{+0.04}$ & 148 $\pm$ 22 & 596 $\pm$  91 & 5.4 $\pm$ 1.6 & $0.09_{- 0.07}^{+ 0.15}$ \\
          COS-26097  & $-20.26_{-0.18}^{+0.18}$ & $-2.56_{-0.14}^{+0.27}$ & $8.6_{-0.5}^{+0.5}$ & $ 6.4_{-1.2}^{+1.9}$ & $0.03_{-0.02}^{+0.07}$ &  82 $\pm$ 17 & 343 $\pm$  71 & 1.0 $\pm$ 0.3 & $0.18_{- 0.16}^{+ 0.11}$ \\
          COS-22493  & $-20.09_{-0.19}^{+0.20}$ & $-2.18_{-0.25}^{+0.40}$ & $9.0_{-0.6}^{+0.4}$ & $ 8.1_{-3.1}^{+3.2}$ & $0.09_{-0.07}^{+0.12}$ &  73 $\pm$ 15 & 315 $\pm$  66 & 1.2 $\pm$ 0.4 & $0.22_{- 0.19}^{+ 0.10}$ \\
          COS-12634  & $-20.08_{-0.24}^{+0.22}$ & $-2.54_{-0.19}^{+0.37}$ & $8.5_{-0.4}^{+0.5}$ & $ 4.4_{-1.6}^{+1.3}$ & $0.04_{-0.03}^{+0.09}$ &  71 $\pm$ 18 & 298 $\pm$  77 & 0.7 $\pm$ 0.3 & $0.16_{- 0.14}^{+ 0.13}$ \\
          COS-14608  & $-20.04_{-0.26}^{+0.34}$ & $-1.97_{-0.34}^{+0.50}$ & $9.3_{-0.7}^{+0.3}$ & $ 3.9_{-2.8}^{+5.3}$ & $0.13_{-0.09}^{+0.15}$ &  78 $\pm$ 21 & 334 $\pm$  91 & 1.6 $\pm$ 0.5 & $0.28_{- 0.11}^{+ 0.09}$ \\
          COS-4396  & $-19.99_{-0.14}^{+0.18}$ & $-2.20_{-0.17}^{+0.27}$ & $8.4_{-0.2}^{+0.4}$ & $ 7.2_{-2.3}^{+1.6}$ & $0.05_{-0.03}^{+0.09}$ &  50 $\pm$ 17 & 212 $\pm$  71 & 0.7 $\pm$ 0.3 & $0.11_{- 0.09}^{+ 0.13}$ \\
          COS-24512  & $-19.94_{-0.15}^{+0.21}$ & $-2.72_{-0.11}^{+0.17}$ & $8.3_{-0.3}^{+0.4}$ & $ 3.1_{-0.5}^{+0.6}$ & $0.01_{-0.01}^{+0.04}$ &  48 $\pm$ 27 & 207 $\pm$ 115 & 0.5 $\pm$ 0.3 & $0.05_{- 0.04}^{+ 0.20}$ \\
          COS-10016  & $-19.85_{-0.18}^{+0.23}$ & $-2.42_{-0.23}^{+0.36}$ & $8.5_{-0.5}^{+0.5}$ & $ 4.1_{-1.0}^{+0.8}$ & $0.04_{-0.03}^{+0.09}$ &  55 $\pm$ 25 & 239 $\pm$ 109 & 1.4 $\pm$ 0.6 & $0.18_{- 0.16}^{+ 0.12}$ \\
          COS-21431  & $-19.47_{-0.24}^{+0.36}$ & $-2.08_{-0.32}^{+0.49}$ & $8.6_{-0.6}^{+0.5}$ & $ 1.5_{-0.9}^{+1.0}$ & $0.11_{-0.08}^{+0.14}$ &  61 $\pm$ 34 & 269 $\pm$ 151 & 0.5 $\pm$ 0.2 & $0.21_{- 0.19}^{+ 0.11}$ \\
          COS-3755  & $-19.46_{-0.22}^{+0.26}$ & $-2.59_{-0.13}^{+0.29}$ & $7.3_{-1.6}^{+1.5}$ & $ 2.2_{-0.7}^{+0.3}$ & $0.02_{-0.02}^{+0.06}$ &  67 $\pm$ 27 & 287 $\pm$ 117 & 0.5 $\pm$ 0.3 & $0.27_{- 0.16}^{+ 0.10}$ \\
\enddata
\tablecomments{(1) ID ordered by M$_{\rm UV}$, (2) UV magnitude, (3) rest-frame UV slopes, (4) Stellar mass, (5) star-formation rate averaged over 10 Myrs, (6) dust attenuation A$_{V}$, (7) and (8) radius in mas and in parcsec at the redshift (\beagle) of the source,  (9) \sersic\ index, and (10) age of galaxy after formation of 50\% of its stellar mass.}
\end{deluxetable*}

\section{Results}
\label{sec:results}

In Table~\ref{tab:redshift}, we summarize the redshifts of our 15 candidates from each of the SED fitting approaches. We also show the goodness of fit based on the $\chi^{2}$ between the \textsc{EAzY} fits when run in full and low$-z$ modes, respectively.

\subsection{The sample}
\label{subsec:sample}
Here, we present our sample of $z\ge9$ galaxies selected in the first COSMOS-Web data. Figure~\ref{Fig::position} shows a color image of the COSMOS field in the four NIRCam bands, with the positions of our 15 $z\ge9$ candidates overlaid. 

In Figure~\ref{Fig::SEDs}, we show the best-fitting UV-optical SEDs of the galaxies, along with their redshift probability distribution functions. We show cutout stamps for each galaxy above the SED, from \textit{HST}/F814W to MIRI F770W in Appendix (see Fig.~\ref{Fig::cutouts}). In all cases, the galaxy emission becomes more prominent towards redder wavelengths, with clear detections in the NIRCam F150W to F444W bands.

\subsection{Properties of $z \ge 9$ galaxy candidates}
\label{subsec:properties}
In this section, we discuss the properties of this sample of $z\ge9$ galaxies. In Table~\ref{tab:properties}, we summarise the stellar masses, UV magnitudes, $\beta$-slopes, star-formation rates (SFRs), dust attenuations A$_{V}$, radii and \sersic\ indices for our sample.

\subsubsection{Redshift}
\label{subsubsec:redshift}
The best-fit photometric redshifts derived by \textsc{EAzY}, \beagle\  and \bagpipes\  are shown in the top-right of each panel in Figure~\ref{Fig::position}, in addition to the redshift probability distribution function (PDF) in the upper left corner of each panel. The redshift of our sample (from \textsc{EAzY}) varies between 9.2 and 10.7. This means that we selected only F115W dropouts. No galaxy in our sample is a F150W dropout.  We find good agreement between the different redshift estimations (see Fig.\ref{Fig::Redshift}). All derived redshifts, regardless of the technique, are in agreement within their uncertainties.  For this sample, we find a difference $\Delta z$ /(1+$z_{\rm mean}$) of 0.02 between \bagpipes\  and \textsc{EAzY}, 0.05 between \beagle\  and \bagpipes\  and 0.02 between \beagle\  and \textsc{EAzY}. \bagpipes\  gives systematically slightly lower redshifts than the two other SED fitting codes.

Without spectroscopic confirmation of these sources, there may be doubts about the reliability of these redshifts. Since the publication of the first studies on LBG candidates at  $z>10$, concerns have been raised that some of these candidates may be low-redshift dusty contaminants \citep[e.g.,][]{zavala23}, which could significantly impact our understanding of early galaxy formation.  While spectroscopic redshifts are now trickling in at $z \sim$ 9--13 \cite[e.g.,][]{arrabal23, arrabal23b, fujimoto23, curtis-lake23, robertson23}, most high-redshift candidates still exhibit secondary redshift solutions at $z \sim$ 3--6.  Recently, \cite{arrabal23} have shown that the galaxy previously claimed to have the highest photometric redshift (CEERS-93316, $z_{\rm phot}$$\sim$16; \citealt{donnan23a}) was, in fact, at $z_{\rm spec}$ = 4.9, with an SED exhibiting the signature of a dusty star-forming galaxy, with strong nebular lines mimicking the Ly$\alpha$ break.  Models often make the assumption that these galaxies have a red color that sets them apart from the typically blue LBGs. The intricate interstellar medium (ISM) environments present in dusty star-forming galaxies (DSFGs; see review by \citealt{casey14}), along with contamination from nebular emission lines, could result in a variety of observed near-IR colors \citep{naidu22,perez-gonzalez23,zavala23,mcKinney23}. However, we used a selection technique relatively similar (but a different number of filters and depth) as in the CEERS field, \citet{finkelstein23a,fujimoto23} achieve a spectroscopic confirmation rate of $\sim$90\% for galaxies with $z\sim8-9$. It is interesting to note that in this latter study, the photometric redshifts (derived with \textsc{EAzY}) are for the most part (6/7) higher (by $\Delta z  \sim$ 1-2) compared to the spectroscopic redshifts that were determined afterwards. This validates the general reliability of the photometric redshift estimates out to $z\sim9$ close to the redshift range of our sample.

\begin{figure*}
\centering
\includegraphics[width=1.\textwidth]{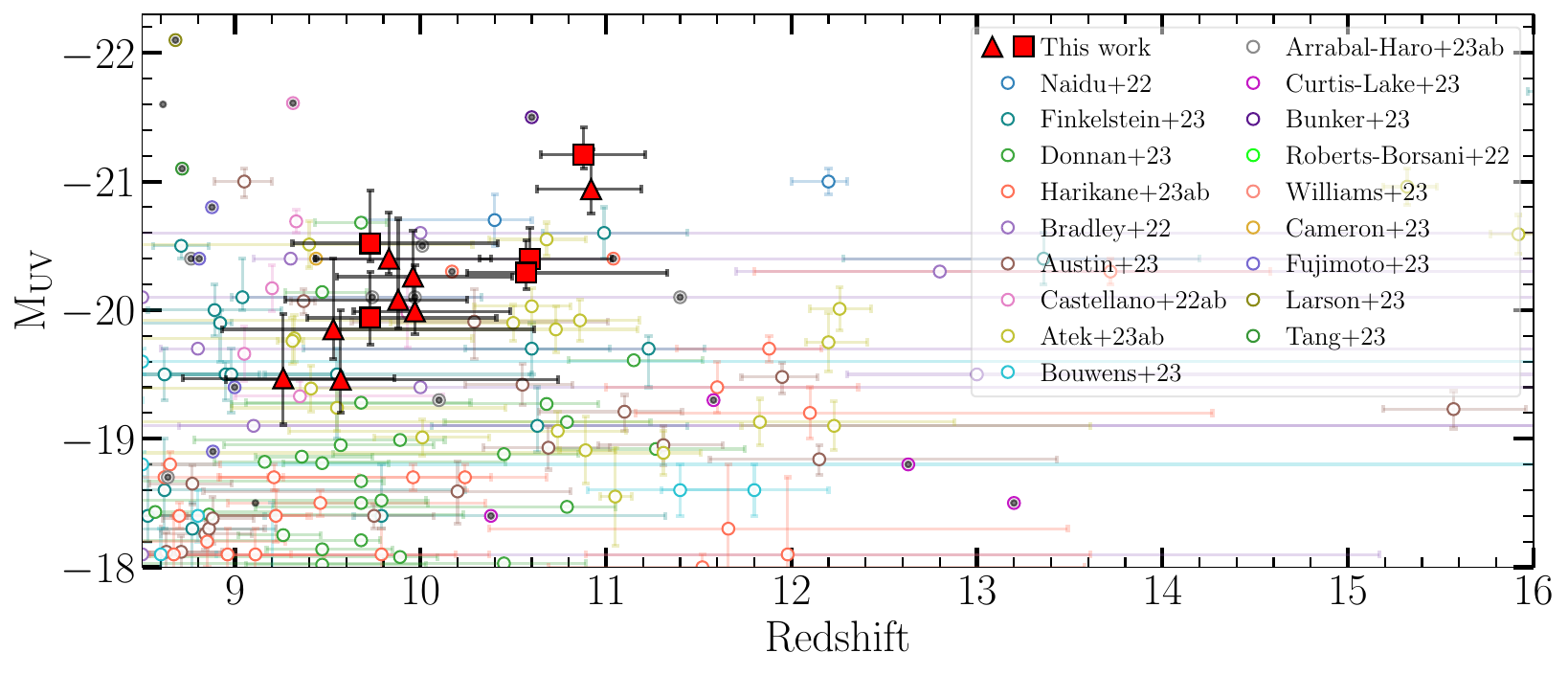}
\caption{Absolute ultraviolet magnitude (M$_{\rm UV}$) as a function of redshift (\beagle) for our z$>$9 galaxy sample (red points). The difference between red squares and red triangles is the same as in Fig.~\ref{Fig::Redshift}. We include data from
literature from \cite{naidu22}, \cite{finkelstein23a}, \cite{donnan23a}, \cite{harikane23}, \cite{harikane23b}, \cite{bradley22}, \cite{austin23}, \cite{castellano22a}, \cite{castellano22b}, \cite{atek23a}, \cite{atek23b}, \citep{bouwens23}, \cite{arrabal23}, \cite{arrabal23b}, \cite{curtis-lake23}, \cite{bunker23}, \cite{roberts_borsani23}, \cite{williams23}, \cite{cameron23},  \cite{fujimoto23}, \cite{larson23}, \cite{tang23}. 
If a galaxy is mentioned in multiple papers and a spectroscopic redshift is available, we have only displayed the one with spectroscopic confirmation. Otherwise, we have taken a conservative approach and displayed the galaxy with the lowest redshift available. Galaxies with spectroscopic confirmation (line or break) are displayed with an additional black dot. At a given redshift, the galaxies in our sample are generally those that display some of the brightest UV magnitudes. }
\label{Fig::M_UV_z}
\end{figure*}

\subsubsection{UV magnitudes and spectral slopes}
\label{subsubsec:MUV}
The rest-frame UV spectrum of a galaxy can be approximated with a power-law of the form $f_\lambda \propto \lambda^\beta$ \citep{calzetti94,meurer99}. Pre-\textit{JWST} studies performed at high redshift ($z>6$) with \textit{HST} have shown that the galaxies presented blue UV slopes with slope values near $-2$ \cite[e.g.,][]{finkelstein12, bouwens14} typical of relatively young and metal-poor galaxies. The main question is to know if at higher redshift, we observe an abrupt break of slope that can reach values of $-$3 as it has been reported recently by \cite{topping22, cullen23, austin23}. We calculated the beta slope of these galaxies by fitting a power law to the best-fit spectrum from \beagle\  between 1268 \AA\  and 2580 \AA\ using the fitting windows given in Table 2 of \cite{calzetti94}. This method has proven its efficiency and accuracy and gives better results than the use of a single color, for example, which is much more subject to photometric outliers. In addition, it is the method with the smallest dispersion when compared with simulations \citep{finkelstein12}. The uncertainties were calculated by generating 800 fits from the posterior and considering the 16 to 84 percentiles. The rest-frame UV slope for our sample ranges between $-$2.0 and $-$2.7 (mean value $-$2.4), with UV magnitudes from $-$19.5 to $-$21.2. To compare the different studies, it is necessary to put them in perspective with the UV magnitude.

\begin{figure*}
\centering
\begin{minipage}[t]{1.\textwidth}
\resizebox{\hsize}{!} { 
\includegraphics[width=0.5\textwidth]{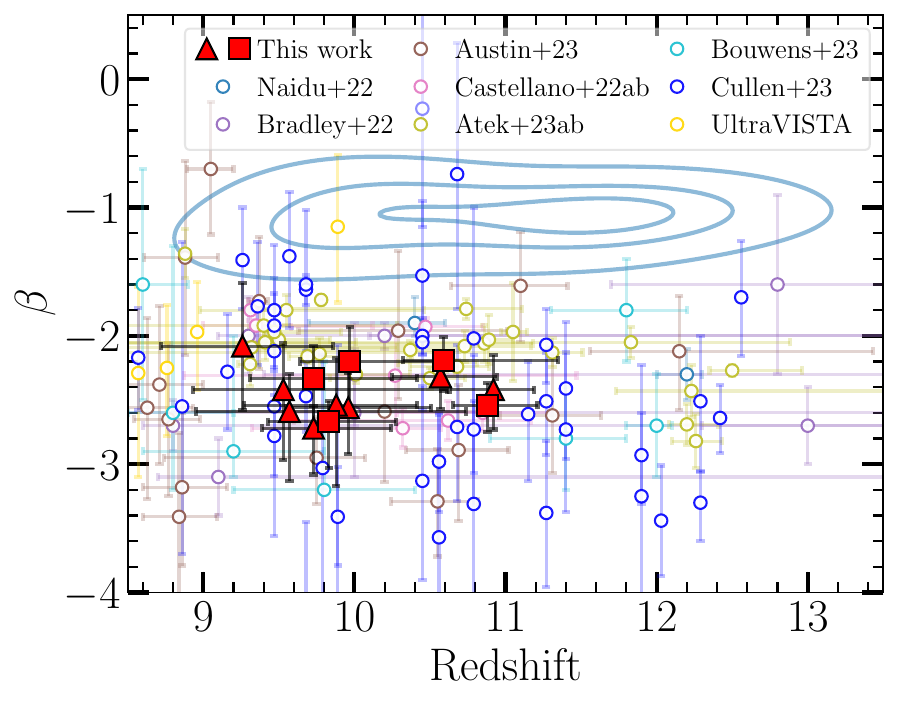}
\includegraphics[width=0.52\textwidth]{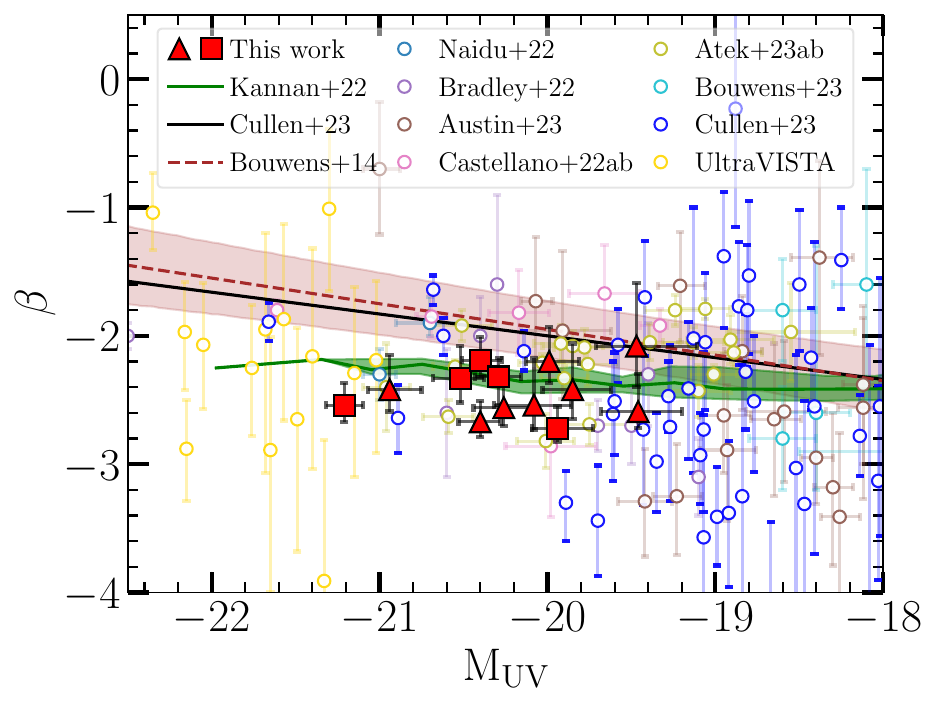}
}
\end{minipage}
      \caption{Left: Ultraviolet spectral slope ($\beta$) as a function of redshift for our $z\ge9$ sample (red points) and a compilation of galaxies detected by \textit{JWST} at $z>8.5$. In addition, we included an extra sample of bright high-redshift galaxies (yellow points) obtained from wide-area ground-based near-IR imaging within the COSMOS/UltraVISTA field (at $z\sim8-10$) from \cite{donnan23a}. The blue contours represent the parameter space occupied by a synthetic sample of galaxies at $5>z>6$, for which their broad emission lines in optical spectra could be important contaminants in F150W dropouts at $z\ge9$ \citep{mcKinney23}. The clear separation of our galaxies from this region provides further support for the accuracy of our redshift measurements. Right: $\beta$ as a function of M$_{\rm UV}$. The best fits from \cite{cullen23} (for $8<z<16$ galaxies) and \cite{bouwens14} (for $z\sim$7 galaxies) and the prediction from the \texttt{THESAN} simulations (for $z\sim9$ galaxies; \citealt{kannan22}) are displayed in black, brown and green respectively. Our galaxies are slightly bluer than the trends presented in \cite{bouwens14} and \cite{cullen23} but they are closely related to the trend from \cite{kannan22}. For each of the two panels, results for our $z\ge9$ galaxy sample are displayed in red and put into perspective with a compilation of recent results from the \textit{JWST} at $z>8.5$. The difference between red squares and red triangles is the same as in Fig.~\ref{Fig::Redshift}. }
         \label{Fig::M_UV_beta}
\end{figure*}

Indeed, many studies point out the evolution of the $\beta$ slope as a function of the UV magnitude \citep[e.g.,][]{bouwens14,cullen23}. This would suggest that the brighter galaxies are also older, more dust-obscured and more metal-enriched than the fainter ones \citep[i.e.,][]{cullen23}. Due to the large contiguous area of the COSMOS-Web survey, it will be possible to explore a wider portion of the parameter space by finding rarer galaxies with brighter UV magnitudes than other studies. 
We calculate the absolute magnitude at 1500 \AA\ using the best-fit spectrum from \beagle. This involves integrating the flux within a 100 \AA\ wide top-hat filter centered on 1500 \AA\ and then conversion to apparent magnitude (m$_{1500}$). We then convert it to an absolute magnitude following:

\begin{equation}
M_{\rm UV}=m_{1500}-5 \log_{10}\left(\frac{D_L}{10}\right)+2.5 \log_{10}(1+z),
\end{equation}
\noindent with $D_L$ the luminosity distance in parsecs. We have verified that the results given by the different SED fitting tools are consistent with each other. The $\beta$ and $M_{\rm UV}$ values are consistent within $\sim$0.2.

In Fig.~\ref{Fig::M_UV_z}, we show the evolution of the absolute ultraviolet magnitude as a function of the redshift for our sample and for a compilation of galaxies detected with \textit{JWST} at $z>8.5$ (\citealt{naidu22}, \citealt{finkelstein23a}, \citealt{donnan23a}, \cite{harikane23}, \citealt{harikane23b}, \citealt{bradley22}, \citealt{austin23}, \citealt{castellano22a}, \citealt{castellano22b}, \citealt{atek23a}, \citealt{atek23b}, \citealt{bouwens23}, \citealt{arrabal23}, \citealt{arrabal23b}, \citealt{curtis-lake23}, \citealt{bunker23}, \citealt{roberts_borsani23}, \citealt{williams23}, \citealt{cameron23},  \citealt{fujimoto23}, \citealt{larson23}, \citealt{tang23}). For the redshift range of our study ($z\sim9-11$), our sample is among the galaxies with the brightest UV magnitude. In particular, the two most distant galaxies in this sample are exceptionally bright (COS-28841 and COS-17810) with M$_{\rm UV}$ = $-21.21^{+0.11}_{-0.11}$ and $-20.94^{+0.19}_{-0.13}$ respectively. These two galaxies are slightly fainter than GN-z11 (M$_{\rm UV}$ = $-\-$21.50$\pm$0.02; \citealt{bunker23}), one of the most luminous galaxy detected at these redshifts.

In Fig.~\ref{Fig::M_UV_beta}-left, we show the evolution of the $\beta$ slope as a function of the redshift. The redshift range of our sample is relatively narrow: $\Delta z\sim1.5$ corresponding to $\sim$100 Myrs. Observing a trend can be challenging.  We did not find any correlation between M$_{\rm UV}$ and $\beta$ in our sample. Our results are slightly bluer than the relationship depicted in Fig.~\ref{Fig::M_UV_beta}-right, which is derived from the work of \cite{cullen23} by $\Delta \beta \approx -0.4$ but well aligned with the predictions from the \texttt{THESAN} project \citep{kannan22a, garaldi22, smith22} simulating the emission line properties of high-redshift galaxies. Further investigation and a larger sample are needed to understand the origin of these somewhat bluer colors.

We note that the breadth of $\beta$ slopes measured in this sample are somewhat narrower than literature samples, though this is most likely due to the difference in approach to measuring $\beta$: we use the best-fit SED to constrain the slope rather than a direct measurement from photometry.  The latter is free from potential bias introduced by the SEDs fit, but introduces other systematics due to the different rest-frame wavelengths of the bands used to calculate $\beta$.

\begin{figure}
\centering
\includegraphics[width=0.5\textwidth]{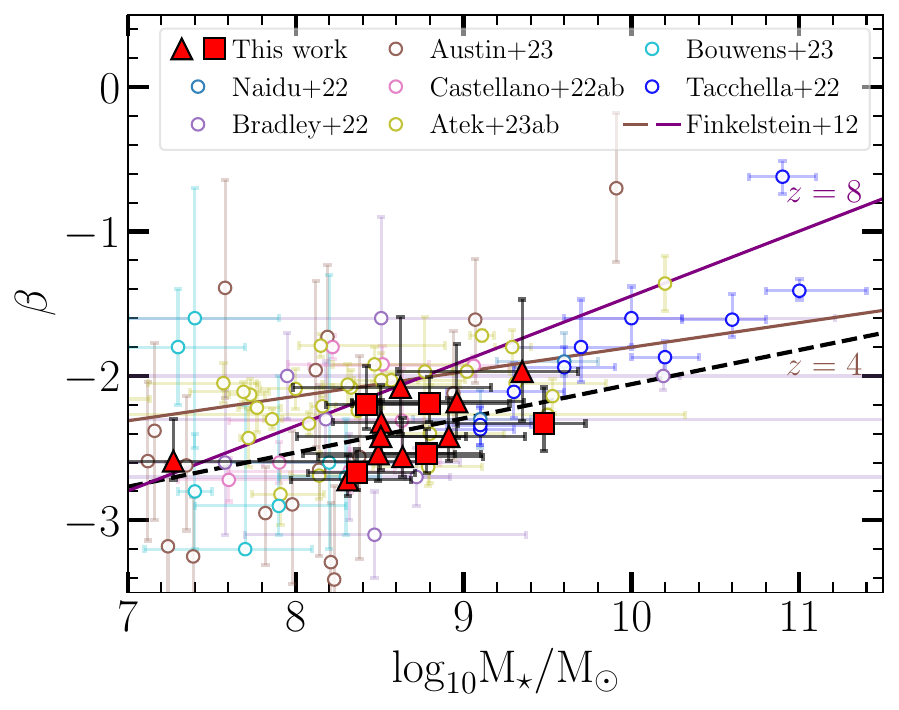}
\caption{Ultraviolet spectral slope ($\beta$) as a function of the stellar mass of our $z\ge9$ sample (red points) and a compilation of galaxies detected by \textit{JWST} at $z>8.5$. In addition, we included an extra sample of bright ($H<26.6$) high-redshift galaxies ($z=8.5-11$) selected by \cite{finkelstein12} in the CANDELS fields \citep{grogin11, koekemoer11} and studied in \cite{tacchella22}. We also added the trend between $\beta$ and stellar mass at $z=4$ and $z=8$ derived by \cite{finkelstein12} (brown and purple lines respectively). While $\beta$ does not show a significant correlation with redshift or absolute UV magnitude, we observe a clear correlation between $\beta$ and stellar mass for our sample, with a slope of 0.24$\pm$0.10 (dashed line).}
\label{Fig::beta_mstar}
\end{figure}

At certain redshifts, dust-obscured galaxies with strong nebular emission lines and high optical attenuation (A$_V$ $>$ 3-5) can mimic the photometry of $z\ge9$ LBGs. In this case, the increase in broadband filter flux by strong optical emission lines can mask the underlying red continuum and give the appearance of a blue UV continuum slope. However, as shown in \cite{mcKinney23}, strong lines are only capable of reproducing $\beta$ up to -1.5 for high-z candidates with $8<z<14$ (blue contours, Fig.~\ref{Fig::M_UV_beta}-left). Our sample falls outside of this confusion regime, which diminishes the likelihood of low redshift DSFGs contaminating our sample in addition to the success rate of e.g., \cite{fujimoto23} as previously discussed.

\begin{figure*}
  \centering
\begin{minipage}[t]{1.\textwidth}
\resizebox{\hsize}{!} { 
    \includegraphics[width=.49\textwidth]{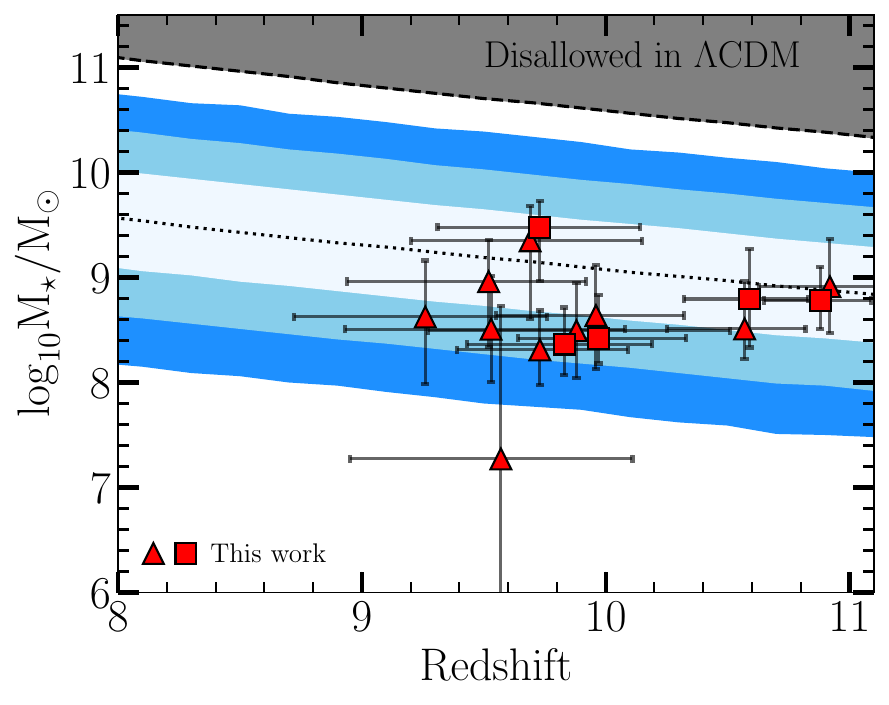}
    \includegraphics[width=0.518\textwidth]{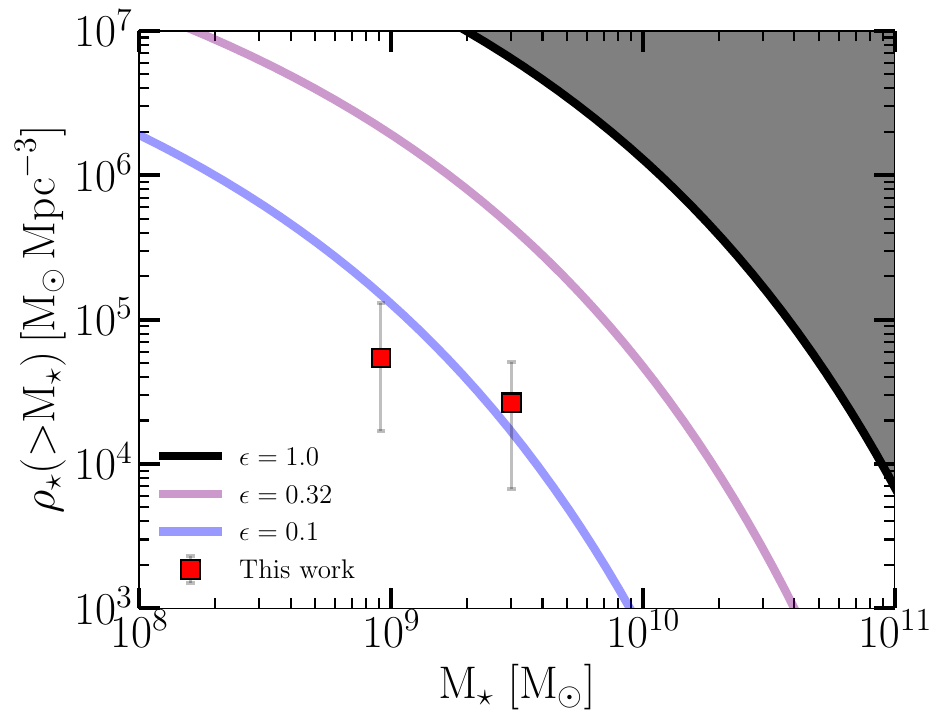}
}
\end{minipage}
  \caption{Left panel: Stellar mass as a function of redshift for our $z\ge9$ sample (red points). The PDF of the most massive galaxy predicted by Extreme Value Statistics (EVS) from \cite{lovell23} is also displayed in shades of blue (1, 2 and 3 $\sigma$) around the median value of the expected maximum stellar mass for the volume of this survey. The dashed line shows the 3$\sigma$ upper limit assuming a stellar fraction of unity. The absence of galaxies more than one sigma above this limit indicates no tension between our observations and the $\Lambda$CDM cosmology. Right panel: The comoving stellar mass density within galaxies that are more massive than M$_\star$ (for two stellar mass bins) at the median redshift of our sample can be expressed for three different assumed values of the conversion efficiency ($\epsilon$) which represents the transformation of a halo's cosmic allotment of baryons into stars. The stellar masses derived in our sample do not present any tension with standard $\Lambda$CDM models. This implies a conversion efficiency $\sim$ 0.1.  The difference between red squares and red triangles is the same as in Fig.~\ref{Fig::Redshift}.} 
  \label{Fig::stellar_mass_}
\end{figure*}

The bluest galaxy in our sample (COS-4393) has an index $\beta=-2.72_{-0.11}^{+0.17}$. Though quite blue, this steepness is not extreme and is even comparable to values derived in the local Universe (e.g, NGC 4861, NGC 1705, Mrk153 with $\beta$ from $-$2.5 to $-$2.4; \citealt{takeuchi12}). This would seem to show that even at z$\sim$10 the stellar population did form from an environment that is not particularly dusty but one that is certainly already enriched in metals. With a mean value of $\beta$=$-$2.4$\pm$0.2, $\beta$ seems to be relatively constant  between $z=7$ and $z=11$ \citep{dunlop13,bouwens14}. 
In contrast to \cite{topping22} at $z=7-8$, or \cite{cullen23} and \cite{austin23} at higher redshifts, we do not find any ultra-blue objects in our sample. This may mean that the mixing of the interstellar medium may be heterogeneous or the properties of the galaxies may be environment dependent or due to a blue-bias in the $\beta$ scatter at faint luminosities (as suggested by \citealt{cullen23}) and only an observation over a much larger field will reveal this. Even if we do not find extremely blue galaxies, we can still note that our sample is systematically slightly bluer at a given UV magnitude than the relations derived by \cite{cullen23} and \cite{bouwens14} but are in good agreement with the prediction from the \texttt{THESAN} simulations for $z\sim9$ galaxies (\citealt{kannan22}; see the right panel of Fig.~\ref{Fig::M_UV_beta}).

While we find no significant dependence between the evolution of the UV spectral slope and the redshift or absolute UV magnitude (Pearson $r$ = -0.13 and 0.18 respectively), we observe a correlation between $\beta$ and stellar mass (Pearson $r > $ 0.5), see Fig.~\ref{Fig::beta_mstar}. We relate these two quantities for our sample by the following equation:
\begin{equation}
    \beta = (0.24\pm0.10)\times \log_{10} (\text{M}_\star/\text{M}_\odot) -4.42\pm0.90
\end{equation}

This means that the more massive a galaxy, the redder its UV spectral slope. We want to emphasize that the level of completeness can strongly influence this scenario, as our dataset tends to detect low-mass blue galaxies more readily than low-mass red galaxies. The effects of completeness on the type of galaxies detected will be studied in detail in a future paper in preparation. However, this correlation between $\beta$ and stellar mass has also been noted in previous studies \citep[e.g.,][]{finkelstein12,tacchella22}.

\subsubsection{Stellar masses and SFRs}
\label{subsubsec:Mstar}

All of the galaxies of our sample lie at relatively low stellar mass, as expected for this high redshift, between log$_{10}($M$_{\star}/$M$_{\odot}$)$\sim$8.3--9.5 with one outlier at log$_{10}($M$_{\star}/$M$_{\odot}$) = 7.3. We also infer low levels of dust obscuration, A$_{V}<$0.2 in all cases, with the majority of galaxies with nearly no attenuation, A$_{V}<$0.1. However, we emphasize that this value is not well constrained directly due to a degeneracy between the redshift and A$_{V}$.

 Fig.~\ref{Fig::stellar_mass_}-left shows the distribution of detected stellar masses as a function of the redshift. We have added to these values the predicted contours from extreme value statistics (EVS; \citealt{gumbel58, kotz00}) derived by \cite{lovell23}, adapted to the surface area of this survey. 
This approach predicts the probability contours of the maximum (or minimum) value of a random variable selected from a given distribution. 
This method has been applied to the halo mass function, in order to derive the most massive halo at a specific redshift \citep{harrison11}, then coupled with a model for the stellar fraction to derive the PDF of the galaxy with the highest stellar mass for a given volume \citep{lovell23}. 
If galaxies were observed to be considerably higher than the anticipated values for the most massive object, it would suggest a conflict with the $\Lambda$ Cold Dark Matter ($\Lambda$CDM) paradigm, or with the astrophysics underlying the stellar to halo mass relation at high redshift. 
In Fig.~\ref{Fig::stellar_mass_}-left panel, the dotted line shows the median of the maximum expected stellar mass for a survey of 77.19 arcmin$^2$ while the shades of blue show confidence intervals at 1, 2 and 3$\sigma$ around this value. 
We assume a baryon fraction of 0.3, and a log-normal distribution of the stellar fraction. The dashed line shows the upper $3\sigma$ limit assuming a stellar fraction of unity.
At a given redshift interval the most massive galaxy in our sample is globally on that median value, with no galaxy more than one sigma above that limit. 
This indicates no tension between our observations and the $\Lambda$CDM cosmology, nor the astrophysics of galaxy evolution at high redshift.

\begin{figure}
  \centering
    \includegraphics[width=0.5\textwidth]{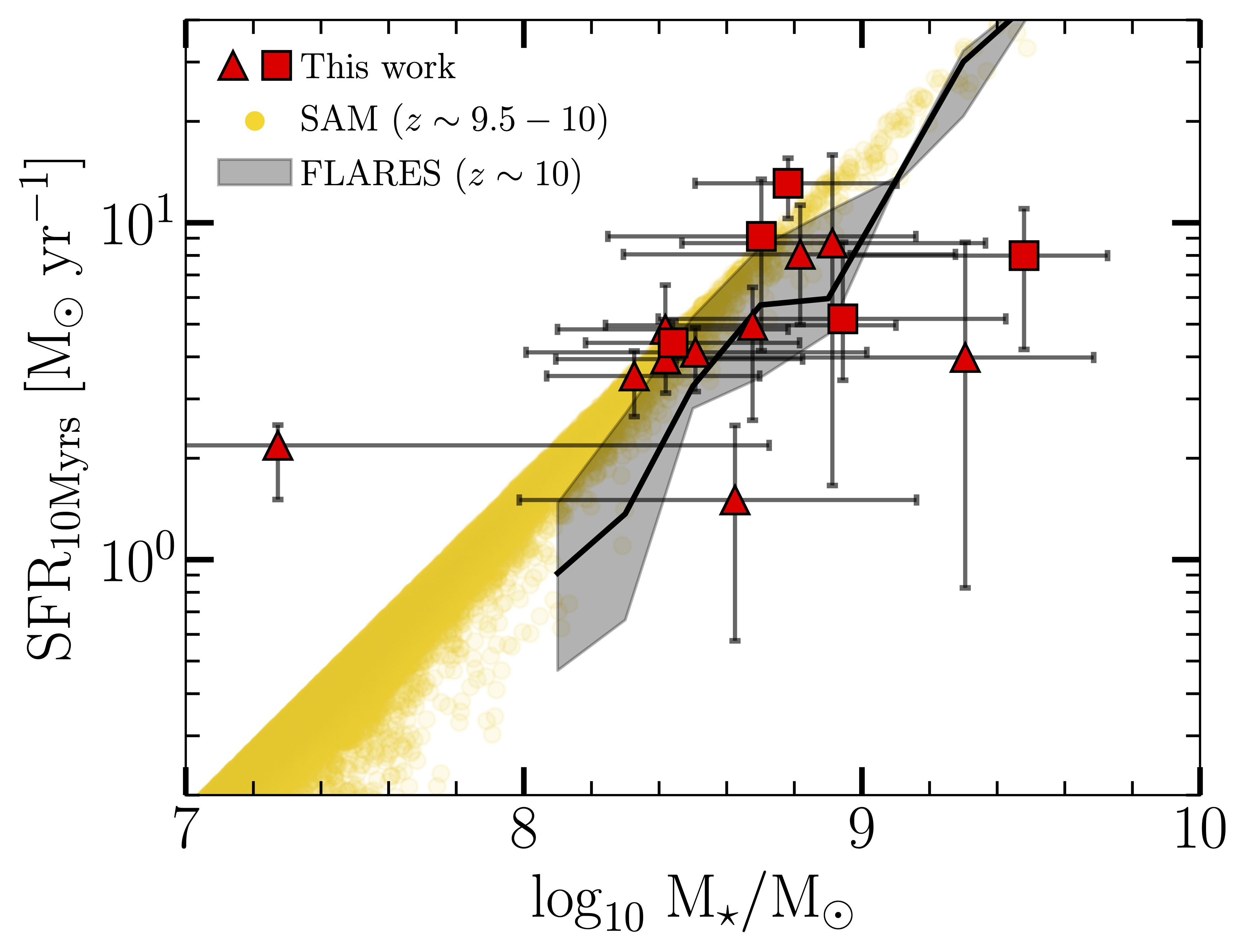}
  \caption{Star formation rate (averaged over 10 Myrs) as a function of the stellar mass for our $z\ge9$ sample (red points). The expected values from the hydrodynamic simulation \textsc{Flares} \citep{lovell20, vijayan20} at $z\sim10$ and Santa Cruz semi-empirical simulation (SAM) \citep[e.g.,][]{yung19} for galaxies at $z = 9.5 - 10$ are displayed in black and yellow respectively. The difference between red squares and red triangles is the same as in Fig.~\ref{Fig::Redshift}.}
  \label{Fig::SFR_Mstar}
\end{figure}

The average SFRs over 10Myrs derived for this sample of galaxies by the Dense-Basis SED fitting code are between 1 and 13 M$_\odot$ yr$^{-1}$. Differences in SFHs from SED fitting codes may cause these values to vary slightly. Although our values show a wide dispersion they are in good agreement with the expected values from the hydrodynamic simulation \textsc{Flares} \citep{lovell20, vijayan20} at $z\sim10$ and Santa Cruz semi-empirical simulation (SAM) \citep[e.g.,][]{yung19} for galaxies at $z = 9.5 - 10$ (Fig.~\ref{Fig::SFR_Mstar}).

\subsubsection{Sizes and morphologies}
\label{subsubsec:Sizes}
\begin{figure}
  \centering
    \includegraphics[width=0.5\textwidth]{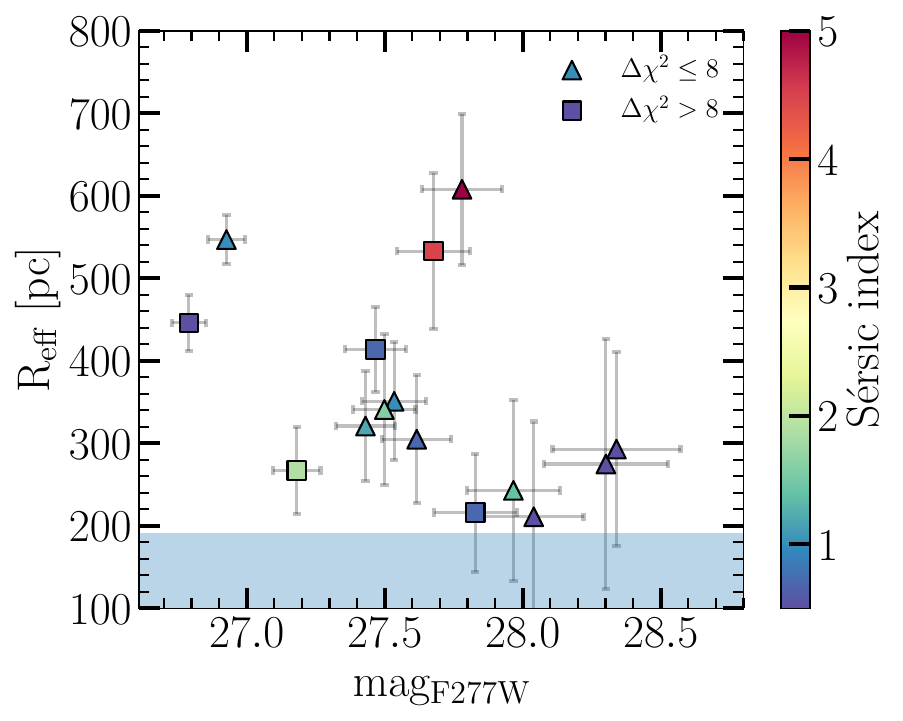}
  \caption{Effective radius as a function of the F277W magnitude for our $z\ge9$ sample measured by \texttt{SE++}, color-coded according to the \sersic\ index. The shaded region represents the effective radius of the F277W \textit{JWST} PSF (0.092 arcsec) at the average redshift of our sample ($z_{\text{mean}, \textsc{beagle}}$ = 9.98). The detected galaxies are compact. While the faintest galaxies are consistent with unresolved sizes, the brightest galaxies are clearly resolved.}
  \label{Fig::size}
\end{figure}

To characterize the rest-frame optical sizes and \sersic\ index of our sample, we utilize \texttt{SE++}. If the measured size uncertainty is less than that derived in Eq.~21 of \cite{condon97}, we rescale the uncertainty according to this equation. Our sample sizes are compact, as none of the galaxies in our sample have an effective radius larger than 0.7 kpc. We find a mean effective radius of 0.37$\pm$0.13\,kpc in F277W which is consistent with the rest-frame UV sizes detected in other surveys such as GLASS \citep{yang22} or CEERS \citep{finkelstein23a} which have median sizes of 0.41 kpc (F277W) and 0.46 kpc (F200W) respectively. In Fig.~\ref{Fig::size}, we show the distribution of sizes as a function of F277W magnitude. Except for the two galaxies with a high \sersic\ index ($n\sim4$), we observe a trend linking magnitude and size, the brightest galaxies being clearly resolved. This correlation could come from the capacity to measure sizes when the S/N is higher but could also be a physical effect. Recent studies by \cite{marshall22} have shown a correlation between galaxy size and far-UV luminosity when measured from dust-attenuated images (this correlation is reversed when the effects of dust are not taken into account) and suggest that dust is the main cause of this correlation. While the two with high \sersic\ indices are measured to be spatially resolved on scales of 600-700\,pc, larger than the median size of the sample, though the uncertainty on their sizes is also higher. In the F277W filter, the PSF is 0.092 arcsec representing an effective radius of 192 pc at the average redshift of our sample ($z_{\text{mean}, \textsc{beagle}}$ = 9.98). 

\section{UV luminosity function}
\label{sec:UVLF}
The rest-frame ultraviolet luminosity function (UVLF) is a crucial observational tracer of early galaxy evolution. The measured volume density of UV-luminous galaxies can be directly compared to simulations to better understand the physical mechanisms that drive galaxy evolution. Observations from the first year of \textit{JWST} observations have now grown large enough to allow the direct calculation of the UVLF beyond pre-\textit{JWST} limits at $z\sim9$.  

To estimate the contribution of our sample, we calculated the volume based on the area covered by the survey and the redshifts and M$_{\rm UV}$ derived by \beagle. The luminosity function point for this sample is computed utilizing the V$_{\rm max}$ method \citep{schmidt68}. The number density of galaxies within a specific magnitude range relies on the maximum volume (V$_{\rm max}$ ) within which each galaxy could have been chosen. The co-moving number density of sources per absolute magnitude, $\Phi\left(M_{\mathrm{UV}}\right)$ is calculated as follows:

\begin{equation}
\Phi\left(M_{\mathrm{UV}}\right) \Delta M_{\mathrm{UV}}=\sum_{i=1}^N\left(\frac{1}{V_{\text{max}, i }}\right),
\end{equation}
\noindent where the volume, for a given galaxy $i$, is computed as:
\begin{equation}
V_{\rm max , i}=\int_{\Omega} \int_{z_{\min , i}}^{z_{\max , i}} \frac{\mathrm{d} V}{\mathrm{~d} \Omega \mathrm{d} z} \mathrm{~d} \Omega \mathrm{d} z,
\end{equation}
\noindent with z$_{\min , i}$ and z$_{\max , i}$ here defined as the 95\% confidence interval for the redshift derived with \beagle. The associated Poissonian uncertainties are given, for the UV luminosity function by: 
\begin{equation}
\sigma_\phi(M_{\mathrm{UV}}) \Delta M_{\mathrm{UV}}=\sqrt{\frac{1}{N} \sum_{i=1}^{N} \frac{1}{V_{\max , i}^2}},
\end{equation}
\noindent where $N$ is the number of galaxies within the UV magnitude range.
In this work, we have not included a completeness correction nor correction for potential contamination, which will be the subject of a future paper with larger samples.

\begin{figure}
  \centering
    \includegraphics[width=0.5\textwidth]{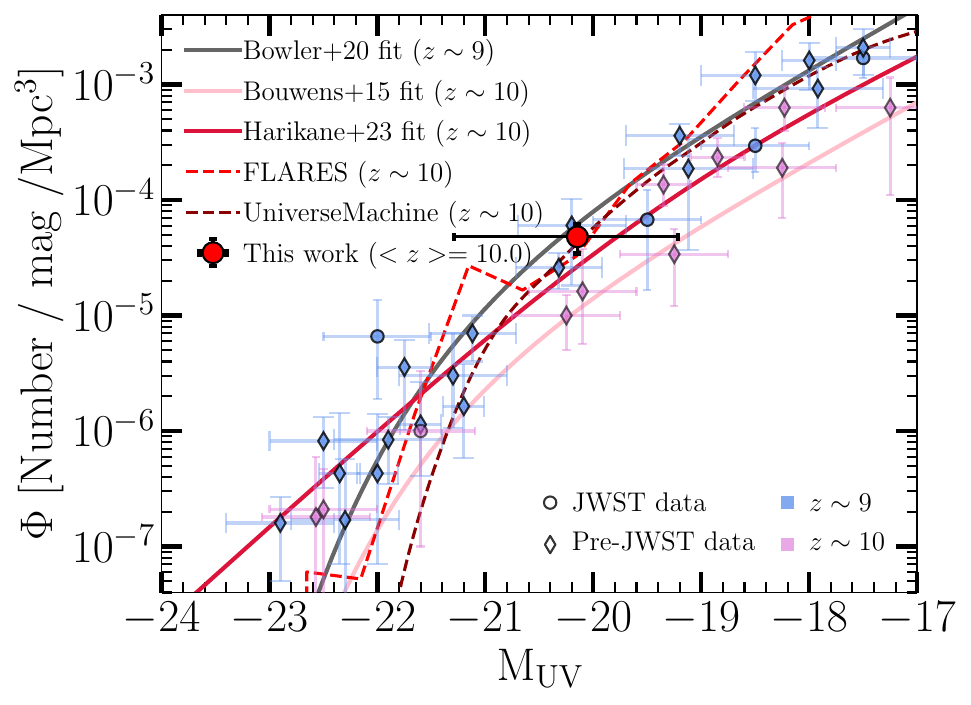}
  \caption{Rest-frame UVLF at $z\sim9$. We include data from literature from \cite{mclure13,oesch13,bouwens13,bouwens15,finkelstein15,finkelstein16,mcleod16,stefanon19,bowler20,bouwens21,kauffmann22,naidu22, donnan23a, harikane22, perez-gonzalez23} at $z\sim$9 and at $z\sim$10 (gray-blue and pink points respectively). The best-fitting Schechter functions at $z\sim$9 from \citep{bowler20} and from \citep{bouwens15} at $z\sim10$ are shown with the gray and red solid lines respectively.  The double power-law fit at $z\sim10$ from \cite{harikane23} is also displayed in red. We differentiate points obtained with the \textit{JWST} (circles) from points obtained before its launch (diamonds). In addition, we also display predictions from the \textsc{Flares}  \citep{lovell20, vijayan20} and the \textsc{UniverseMachine} \citep{behroozi19} simulations at $z\sim10$.} 
  \label{Fig::UVLF}
\end{figure}

 In Fig.~\ref{Fig::UVLF} we present a compilation of several studies at z$\sim$9 from \cite{mclure13,oesch13,bouwens13,bouwens15,finkelstein15,finkelstein16,mcleod16,stefanon19,bowler20,bouwens21,kauffmann22,naidu22, donnan23a, harikane22, perez-gonzalez23, harikane23} as well as the best-fitting Schechter functions from \citep{bouwens15} and \cite{bowler20} and the double power-law function from \cite{harikane23}.  At the mean redshift of our sample ($z=10.0$), we measure a volume density of (4.81$\pm$1.35)$\times10^{-5}$\,Mpc$^{-3}$ per magnitude at M$_{\rm UV}$ = $-$20.19$_{-1.1}^{+0.9}$.  This is about a factor of 1.9 times above expectation from the $z\sim10$ functional form of the UVLF derived by \cite{harikane23} but well aligned with some other works that find a relative excess of $z>10$ candidates \citep{finkelstein23a}. We also compared our results with predictions from the hydrodynamic simulations of galaxy formation and evolution \textsc{Flares} \citep{lovell20, vijayan20} and the \textsc{UniverseMachine} \citep{behroozi19} simulations at $z\sim10$. Our results are in very good agreement with these simulations. As discussed previously in Section~\ref{subsubsec:Mstar}, despite this relative excess of UV bright systems, none of these galaxies exceeds allowable expectations for galaxy formation at these redshifts within a $\Lambda$CDM framework.

\section{Discussion}
\label{sec:discussion}

Interestingly, early \textit{JWST} measurements of the UVLF show an abundance of galaxies which is evolving more shallowly downward with increasing redshift than predicted by simulations \citep[e.g.,][]{harikane22, finkelstein23a, harikane23b}. This result may indicate that at $z\ge9$, the global star formation efficiency (i.e., fraction of baryons in a halo converted into stars) may be higher than at lower redshifts, or/and the stellar initial mass function (IMF) may be top-heavy. Either of these would result in galaxies being more UV luminous than predicted, leading to the observed excess. Both of these may be expected in the first 500 Myr of cosmic time at $z\ge9$, when the physical conditions present in star-forming regions are vastly different from today.  COSMOS-Web provides access to a specific range of parameters for the epoch of reionization that cannot be explored by smaller surveys focusing on only the brightest galaxies (with M$_{\rm UV} \leq -20$), as noted by \cite{casey23a} and \cite{finkelstein23a}. This parameter space is crucial for determining the upper limit of the UV luminosity function and identifying any potential overabundance of bright galaxies during the EoR.  This study will be conducted in detail in a future paper and is outside the scope of this paper.

The anticipated count of galaxies at $z\ge9$ in COSMOS-Web, derived through a direct calculation based on the gathered UV luminosity functions is 8-10 \citep{casey23a}. As such, we have detected 50\% more sources than the upper limit of these predictions. This is intriguing and may have several causes. In contrast to pre-JWST studies, we benefit from unprecedented near-infrared resolution and sensitivity at $\lambda>1.6\mu$m. In particular, in this study, the detection of very high redshift sources is made possible by the long wavelength filters, F277W and F444W. It could also be caused by the cosmic variance. Star formation in the $z>7$  Universe is expected to be highly clustered, with $\sim$40-50\% of the star formation rate density concentrated in the progenitors of massive galaxy clusters \citep[e.g.,][]{chiang17}. 

This could come from an excess of bright galaxies and an underestimation of the bright-end of the luminosity function. This excess of bright galaxies starts to be visible for galaxies both very bright (M$_{\rm UV} \sim -22$) detected with \textit{HST} \citep[e.g.,][]{bagley22a} in contrast to the predicted smooth evolution of a Schechter function from lower redshifts. This excess of galaxies is also observed with the \textit{JWST} at lower UV absolute magnitudes \citep[M$_{\rm UV} \sim -19 -20$, e.g.,][]{finkelstein23a}. A last possibility would be a contamination by low redshift galaxies. However, in view of the arguments presented in Section~\ref{subsubsec:redshift}, we consider that this is not the hypothesis to be favored here. We will explore these possibilities in detail in a future paper. We would point out, however, that we have not taken into account the effects of completeness and contamination in calculating the UVLF. This will be the subject of a future paper. This would tend to qualify our point as a lower limit in the UVLF rather than a measurement.

Recent studies have evoked the hypothesis that galaxies detected by the \textit{JWST} for an area half the size as the area covered by the first part of the COSMOS-Web survey could have masses so high that they were difficult to realize in a standard $\Lambda$CDM cosmology \citep[e.g.,][]{labbe23} at $7.4<z<9.1$. The galaxies presented in this study have a much lower stellar mass with stellar masses derived with the dense-basis SED fitting code ranging between 1.8$\times$10$^{7}$M$_\odot$ and 3.0$\times$10$^{9}$M$_\odot$. Stellar mass values using different SFHs and different SED fitting codes are consistent with these results. We have compared these masses with the method of extreme value statistics derived by \cite{lovell23} and adapted to the solid angle of this survey. It would have been necessary to detect galaxies about $\sim$50 times more massive (assuming a baryon to stellar conversion rate of 1) to enter a parameter space disallowed by $\Lambda$CDM for the most massive galaxies. We have also used the study of \cite{boylan-kolchin23} and adapted it to our case to compare the cumulative mass of our sample with the available supply of baryonic matter within dark matter haloes. We are well below the theoretical value if all available baryons had been converted into stars (i.e., with an efficiency of converting baryons into stars - $\epsilon$ - equal to unity) and the cumulative mass of galaxies detected in this study. For our sample, we find a value of $\epsilon$ about $\sim$ 0.1 (see Fig.~\ref{Fig::stellar_mass_}-right panel). We do however note that at high redshift, the stellar masses can be uncertain due to an evolution of the IMF at very high redshift, when the interstellar medium is less rich in metals, and to the contribution of AGN that complicates stellar mass estimates \cite[e.g.,][]{labbe23}.

\section{Conclusions}
\label{sec:conclusions}
In this paper, we report the detection of high-redshift candidates with redshifts greater than 9 using the initial \textit{JWST} release of COSMOS-Web. These observations cover 77 arcmin$^2$ with the four NIRCam filters (F115W, F150W, F277W, F444W) with an overlap with MIRI (F770W) of 8.7 arcmin$^2$.

We detect 15 galaxies within the redshift range (9.3-10.9) representing a period, relatively short in duration, between 400 Myr and 500 Myr after the Big Bang, a time of rapid change in galaxy evolution.

We have used three SED fitting codes to derive the redshift of these galaxies (\textsc{EAzY}, \beagle, \bagpipes). These three codes are in good agreement with redshifts agreeing between them within their uncertainties. However, we can note that \bagpipes\  systematically gives slightly lower redshifts than the two other SED fitting codes. Only spectroscopic follow-up of these sources will allow us to determine in a robust way the redshifts of these galaxies.

We divided our sample in two parts in oder  to separate galaxies with more robust redshift ($\Delta \chi^2 > 8$) from those with less robust redshifts ($\Delta \chi^2<8$) where $\Delta \chi^2$ represents the difference between the \textsc{EAzY} fit without any constraints ($0<z<15$) and the ``low''-redshift solution ($z<7$). We did not find any significant difference for all the parameters presented in this paper between these two sub-samples which would suggest that our derivation of redshifts is robust.

Although the galaxies detected in this study all have blue UV slopes ($-2.7<\beta<-2.0$), we did not detect any extreme values ($\beta < -$3) as has been reported recently in other studies at similar redshifts. On average, these galaxies display greater luminosities in comparison to the majority of $z\sim9$ candidates discovered by \textit{JWST}, as documented in existing literature. Surprisingly, despite their increased brightness, these galaxies exhibit similar blue hues in their rest-frame UV colors. This observation suggests that, even at 400Myr after the Big Bang, the star-formation occurs from an ISM that is already enriched in metals.

We have derived the UV luminosity function for our sample. We measure a volume density of (4.81$\pm$1.35)$\times10^{-5}$\,Mpc$^{-3}$ per magnitude at M$_{\rm UV}$ = $-$20.19$_{-1.1}^{+0.9}$. We find an excess of bright galaxies with a detection almost twice the expected number of galaxies at $z\ge9$ in the volume of this survey. We therefore find a value above the fit derived by \cite{harikane23} at $z\sim10$, or \cite{bouwens15} prior to the commissioning of \textit{JWST} for galaxies at $z\sim10$ but well aligned with some other works that find a relative excess of $z>10$ candidates \citep[e.g.,][]{finkelstein23a, mcLeod23, adams23b, donnan23a, donnan23b} as well as with predictions from simulations (e.g., \textsc{Flares} or \textsc{UniverseMachine}). This point should be considered only as a lower limit because we do not take into account the effect of completeness.

We have derived the stellar masses of our sample, which are between 1.8$\times$10$^{7}$M$_\odot$ and 3.0$\times$10$^{9}$M$_\odot$. Comparing these results to the maximum expected mass according to $\Lambda$CDM \citep{lovell23,boylan-kolchin23}, we find no tension between observations and theory.

This study focuses on the first observations of the COSMOS-Web survey and represents only 4\% of the total survey area. When the entire survey is completed, the statistics and results given by this study will be refined, the cosmic variance reduced to its minimum and the statistics will be sufficient to have robust constraints on the bright end of the UVLF during the EoR. These results will be complementary with other deeper but smaller \textit{JWST} surveys probing different regions of parameter space.

\begin{acknowledgments}
We thank Sandro Tacchella and the NIRCam team for sharing the wisps templates. We acknowledge that the location where most of this work took place, the University of Texas at Austin, sits on the Indigenous lands of Turtle Island, the ancestral name for what now is called North America. Moreover, we would like to acknowledge the Alabama-Coushatta, Caddo, Carrizo/Comecrudo, Coahuiltecan, Comanche, Kickapoo, Lipan Apache, Tonkawa and Ysleta Del Sur Pueblo, and all the American Indian and Indigenous Peoples and communities who have been or have become a part of these lands and territories in Texas. MF thanks Gene Leung for helpful discussions. MF acknowledges support from NSF grant AST-2009577 and NASA JWST GO program 1727. Support for this work was provided by NASA through grant JWST-GO-01727 and HST-AR-15802 awarded by the Space Telescope Science Institute, which is operated by the Association of Universities for Research in Astronomy, Inc., under NASA contract NAS 5-26555. CMC thanks the National Science Foundation for support through grants AST-1814034 and AST-2009577 as well as the University of Texas at Austin College of Natural Sciences for support; CMC also acknowledges support from the Research Corporation for Science Advancement from a 2019 Cottrell Scholar Award sponsored by IF/THEN, an initiative of Lyda Hill Philanthropies. The french part of the COSMOS team is partly supported by the Centre National d'Etudes Spatiales (CNES). OI acknowledges the funding of the French Agence Nationale de la Recherche for the project iMAGE (grant ANR-22-CE31-0007). This work was made possible by utilizing the CANDIDE cluster at the Institut d’Astrophysique de Paris, which was funded through grants from the PNCG, CNES, DIM-ACAV, and the Cosmic Dawn Center and maintained by S. Rouberol.

\end{acknowledgments}

\appendix

\section{Cutouts of the $z\ge9$ sample}

\begin{figure*}
\centering
\includegraphics[width=0.9\textwidth]{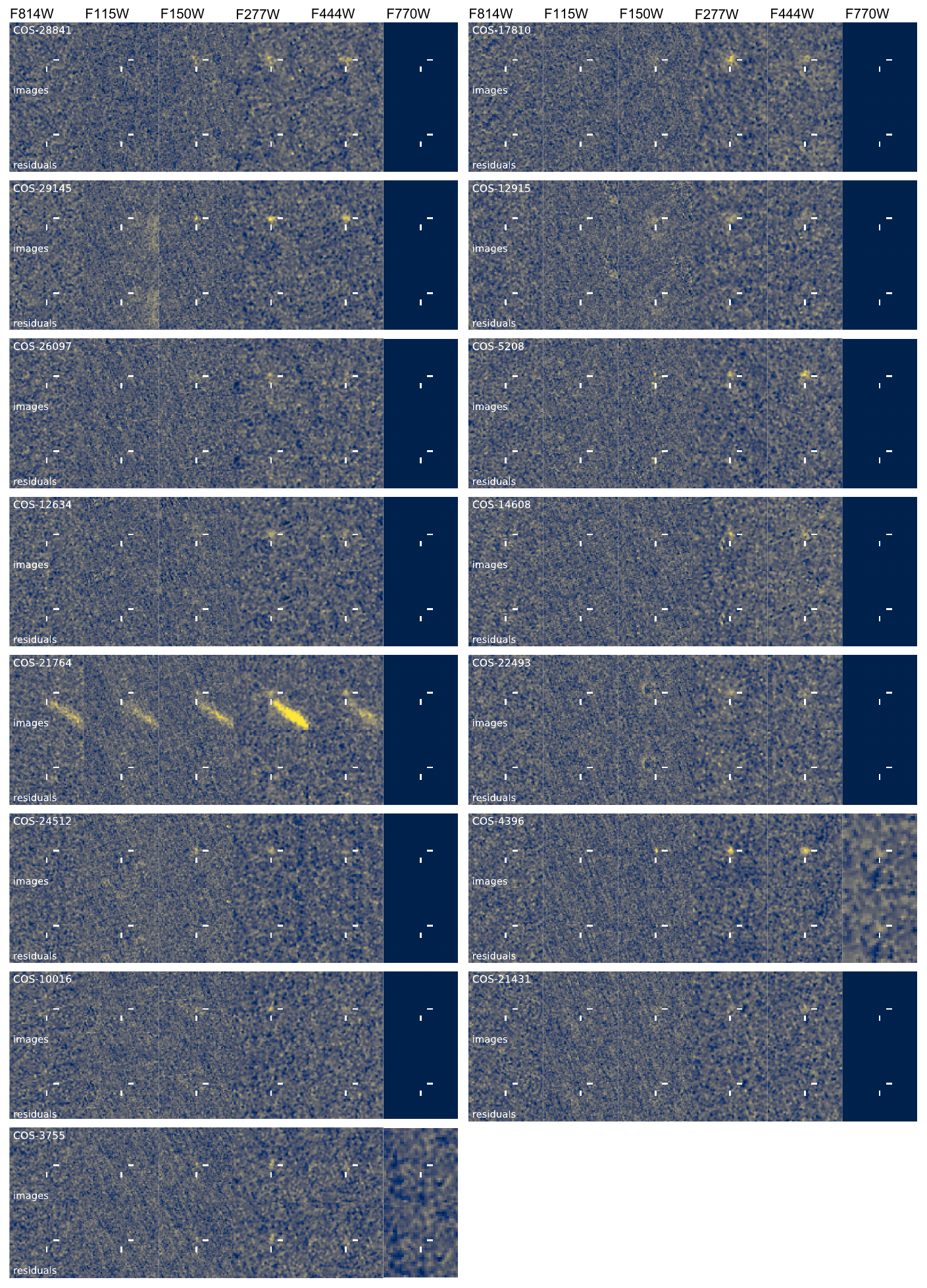} 
      \caption{For each galaxy, we show a 2" stamp image centered on our detections on the upper line and the residuals on the lower line after subtraction of the galaxy models found by \texttt{SE++} for the four NIRCam filters, the MIRI filter and the \textit{HST}/F814W filter. Black squares for MIRI indicate that the galaxy is out of the MIRI coverage.}
       \label{Fig::cutouts}
\end{figure*}

\newpage

\bibliography{franco_cw_gt9}{}
\bibliographystyle{aasjournal}



\newpage

\allauthors

\end{document}